\documentclass[10pt, aps, prd, nofootinbib,
preprintnumbers, showkeys,
superscriptaddress, twocolumn]{revtex4-1}

\usepackage{amsmath, amssymb}
\usepackage{bm}
\usepackage{dcolumn}
\usepackage{siunitx}
\usepackage{graphicx}
\usepackage[caption=false]{subfig}
\usepackage{tikz}
\usepackage{}
\usetikzlibrary{shapes.misc, decorations.markings}

\tikzset{
    cross/.pic = {
    \draw[rotate = 45, very thick] (-#1,0) -- (#1,0);
    \draw[rotate = 45, very thick] (0,-#1) -- (0, #1);
    }
}

\usepackage{booktabs}
\usepackage{multirow}
\usepackage{threeparttable}
\usepackage[normalem]{ulem}
\usepackage{wasysym}
\usepackage{xcolor}

\definecolor{blue}{rgb}{0.0, 0.0, 1.0}
\definecolor{teal}{rgb}{0.0, 128.0, 128.0}
\definecolor{purple}{rgb}{75.0, 0.0, 230.0}
\definecolor{}{rgb}{0.0, 0.14, 0.4}

\def\orcid#1{\kern .08em\href{https://orcid.org/#1}{\includegraphics[keepaspectratio,width=0.7em]{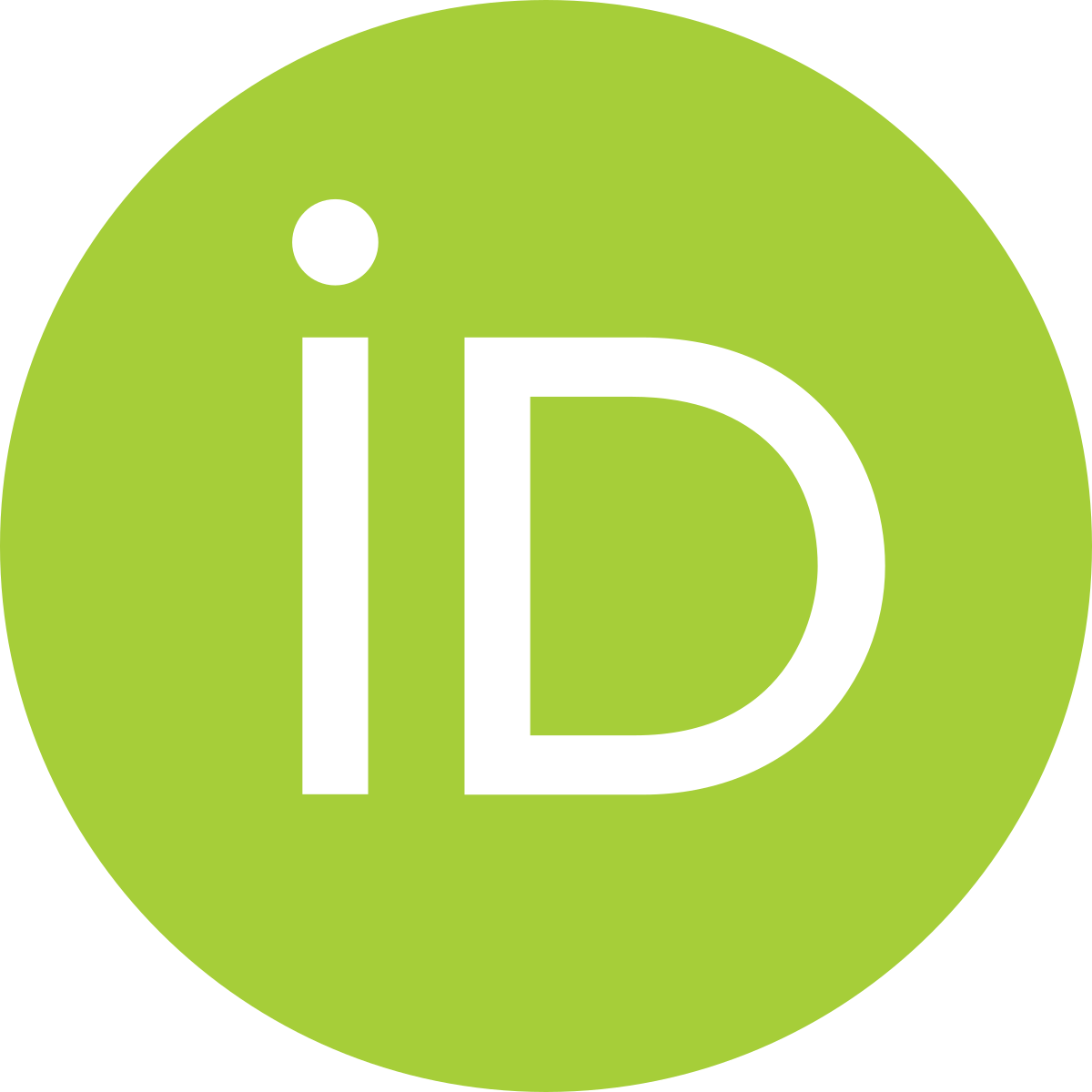}}}

\definecolor{ttcolor}{RGB}{240,248,255}
\definecolor{btcolor}{RGB}{144,238,144}
\definecolor{tbcolor}{RGB}{216,191,216}
\definecolor{bbcolor}{RGB}{255,228,225}

\usepackage{hyperref}
\hypersetup{colorlinks=true, linkcolor=blue, citecolor=purple, urlcolor=blue}

\begin{document}
\title{Deep learning topological inference-guided \texorpdfstring{$T_{cc}^{+}$}{} pole parameter extraction}

\author{Julius B. Pagayon\orcid{0009-0009-2800-2570}}
\email[]{jbpagayon@up.edu.ph}
\author{Klarence Tomas R. Cervantes\orcid{0009-0002-5707-1339}}
\email[]{krcervantes1@up.edu.ph}
\author{Denny Lane B. Sombillo\orcid{0000-0001-9357-7236}}
\email[]{dbsombillo@up.edu.ph}
\affiliation{National Institute of Physics, University of the Philippines Diliman, Quezon City 1101, Philippines}
\date{\today}
\begin{abstract}
We perform a data-driven study of the doubly charmed tetraquark candidate $T_{cc}^+$.  An ensemble of deep neural network classifiers, trained on synthetic amplitudes with controlled analytic structures, identifies a dominant pole topology characterized by an isolated pole on the $[bt]$ Riemann sheet which is robust against left-hand cut effects. A subsequent pole parameter extraction was performed via the uniformized $\mathcal{S}$-matrix and a complementary $\mathcal{K}$-matrix parameterization, which respectively provides a model-independent baseline and dynamical insight on the pole position and trajectory of the resonant state. Using this two-pronged approach, we submit that the $T_{cc}^{+}$ is a shallow $D^0D^{*+}$ bound state in the second Riemann sheet of the complex plane. 
  
\end{abstract}


%%%%%%%%%%%%%%%%%%%%%%%%%%%%%%%%%%%%%%%%%%%%%%%%%%%%%%%%%
%%%%%%%%%%%%%%%%%%%%%%%%%%%%%%%%%%%%%%%%%%%%%%%%%%%%%%%%%

\maketitle

\section{Introduction}\label{sec:intro}

The observation of the doubly charmed tetraquark candidate $T_{cc}(3875)^{+}$ by the LHCb Collaboration in $2021$~\cite{LHCb:2021vvq, LHCb:2021auc} marked a significant milestone in exotic hadron spectroscopy. This exotic state, with quark minimal content $cc\bar{u}\bar{d}$, manifests as a narrow enhancement slightly below the $D^{0}D^{*+}$ threshold in the $D^{0}D^{0}\pi^{+}$ invariant mass spectrum. Despite intensive theoretical efforts, the precise nature of $T_{cc}^{+}$ remains under debate. Various studies propose it to be a compact tetraquark~\cite{Ader:1981db, Heller:1986bt, Carlson:1988ct, Silvestre-Brac:1993ct,Semay:1994ct, Pepin:1997ct, Ballot:1983ct, Zouzou:1986ct, Guo:2021ct, Weng:2021hje, Kim:2022mpa, Dong:2024upa}, a loosely bound $DD^{*}$ hadronic molecule~\cite{Ling:2022hm, Dong:2021hm, Feijoo:2021hm, Fleming:2021hm, Ren:2022hm, Chen:2022hm, Chen:2021vhg, Albaladejo:2022hm, Du:2021zzh, Baru:2022hm, Santowsky:2022hm, Deng:2022hm, Agaev:2022hm, Kamiya:2022hm, Abreu:2022hm, Chen:2022hmPLB, Albaladejo:2022hmEPJ, Peng:2021hm, Ortega:2022hm, Li:2023wug, Dai:2023kwv, Abolnikov:2024key}, or a virtual state~\cite{Dai:2021wxi, Whyte:2024ihh}. An alternative exotic explanation invoke purely kinematical effect (e.g., triangle singularity) but such effects are numerically suppressed~\cite{Achasov:2022onn} in the $T_{cc}^+ \to D^0D^0\pi^+$ decay. One may refer to Refs.~\cite{Chen:2022asf, Francis:2024fwf, Liu:2024uxn} for comprehensive reviews.

In this work, we utilized a two-pronged approach grounded in scattering theory. First, we adopt the coupled-channel $\mathcal{S}$-matrix formalism, constrained by analyticity, unitarity, and hermiticity. To manage the multi-sheeted Riemann surface arising from channel thresholds, we implement a conformal uniformization scheme, which maps the four-sheeted complex $s$-plane into a single-sheeted complex $\omega$-plane~\cite{Kato:1965}. This enables a concise formulation of the amplitude with explicit analytic control over the placement of poles across different Riemann sheets~\cite{Santos:2023gfh}. The uniformization scheme provides significant advantages in parameterization, numerical stability, and interpretability, especially when dealing with near-threshold phenomena where sheet ambiguities are critical to physical interpretation.

Within this framework, we generated a large ensemble of line shapes with known pole topologies and train an ensemble of deep neural network (DNN) classifiers to discriminate synthetic amplitudes into coarse-grained categories based on the number and location of nearby poles. This data-driven method, which builds upon earlier works in single-channel analysis of nucleon-nucleon scattering~\cite{Sombillo:2020ccg, Sombillo:2021ifs} and two-channel analysis~\cite{Sombillo:2021rxv}, has been successfully applied to the study of other resonance phenomena and other application in hadron spectroscopy~\cite{Zhang:2023czx, Santos:2024bqr, Co:2024bfl, Co:2024szz, Frohnert:2025usi, Ng:2024ML, Sadasivan:2025ML, Mapa:2025ML, Dersy:2023ML, Ng:2021prdML, Binosi:2022ML, Chavez:2025ML, Landay:2018ML, Molina:2017ML, Alghamdi:2023ML, Bydzovsky:2021ML, Petrellis:2022ML, Malekhosseini:2024ML}. Here, we extend the method to analyze the experimental $D^{0}D^{0}\pi^{+}$ line shape associated with the $T_{cc}^{+}$, and extract the most probable pole configuration. 

A subsequent pole parameter extraction was performed via the uniformized $\mathcal{S}$-matrix parameterization which provides a model-independent baseline. This was complemented by a dynamical fit to the experimental data using the $\mathcal{K}$-matrix formalism in a coupled-channel setting. The latter enables the extraction of pole positions and tracing of pole trajectory, thereby providing dynamical insight on the underlying nature of the resonant state.

The primary advantage of using DNNs in our analysis lies in their ability to efficiently reduce the vast model space of $35$ pole topology candidates to the most probable one for a detailed analysis. Without this model space reduction process, exhaustive parametric fitting of all pole topologies with the empirical data would be computationally exhaustive and is susceptible to subjective biases. Moreover, unlike end-to-end deep learning approaches that often treat the data as a black box, our hybrid strategy retains a strong connection to underlying physical principles while leveraging the pattern-recognition power of DNNs.

The investigative framework of this work is structured as follows. Section~\ref{sec:2} discusses the coupled-channel scattering framework and the associated uniformization mapping. Section~\ref{sec:3} details the construction of DNN classifiers, including the generation of training dataset, DNN architecture, and the inference to the LHCb data. In Section~\ref{sec:4}, we extracted pole parameters via the uniformized $\mathcal{S}$-matrix and coupled-channel $\mathcal{K}$-matrix model. Section~\ref{sec:conclusion} summarizes our key findings and outlines possible directions for future work.

\section{Independent \texorpdfstring{$\mathcal{S}$}{}-matrix poles} \label{sec:2}

In scattering theory, the scattering amplitude $T(s)$, expressed as a function of the Mandelstam variable $s$ (square of the total center-of-mass (COM) energy $E$), exhibits an analytic structure in the complex $s$-plane that reflects the fundamental properties of the underlying interaction dynamics. Apart from a set of isolated and well-defined singularities, the amplitude is analytic across the complex $s$-plane. These singularities, which include branch points and poles, govern much of the physical content of the amplitude~\cite{Eden:1966, Taylor, Newton:1982}.

Branch points arise whenever a new scattering channel becomes kinematically accessible. This occurs when $\sqrt{s}$ exceeds the threshold energy $\varepsilon_{k} = m_{k1} + m_{k2}$, where $m_{ki}$ is the mass of the $i^{\text{th}}$ particle in channel $k$. Each threshold introduces a square-root-type branch point, indicating that the function is multi-valued. This multivaluedness requires the introduction of multiple Riemann sheets in the complex $s$-plane, with each additional channel doubling the number of sheets. These branch points are accompanied by branch cuts, which are usually defined to extend from the threshold value to positive infinity along the real axis. These cuts are commonly referred to as right-hand cuts, kinematical cuts, or unitarity cuts~\cite{Eden:1966}.

In addition to these kinematical singularities, the amplitude possesses dynamical singularities known as left-hand cuts (lhcs). These arise from the explicit form of the interaction potential, such as the exchange of virtual particles in the $t$ or $u$ channel, and are not related to production thresholds. Their location is determined by the masses of the exchanged particles, typically appearing for $s<0$ or at unphysical values of $s$ along the negative real axis~\cite{Frazer:1964, Oller2019}. While right-hand cuts are universal, the structure of the lhcs is model-dependent and encodes the forces responsible for the scattering, such as the one-pion exchange (OPE) crucial in hadronic interactions.

Moreover, the scattering amplitude also contains pole singularities, which are associated with physical phenomena such as bound states, virtual states, and resonances~\cite{Frazer:1964}. Bound states and virtual states appear on the real $s$-axis below the threshold energy. Bound state poles reside on the physical sheet, while virtual state poles are located on the unphysical sheet. In contrast, resonance poles are located off the real axis and are found in the complex $s$-plane of the unphysical sheets. The analytic nature of $T(s)$ imposes the Schwarz reflection principle, which requires that if a pole exists at a complex point $s_p$, then a corresponding pole must also exist at its complex conjugate $s_p^*$~\cite{Taylor}.

Although the single-channel framework captures many essential analytic features, it becomes insufficient when handling the structure of the $T_{cc}^{+}$ state. The proximity of multiple thresholds and the interplay between open channels necessitates a coupled-channel treatment to accurately describe the pole dynamics.
\subsection{Coupled-Channel Analytic Structure}

In a two-channel system such as the $T_{cc}^{+}$ with dominant open channels $D^0 D^0 \pi^+$ and $D^0 D^{*+}$, the analytic structure is governed by the behavior of the COM momenta for the $k^{\text{th}}$ channel as
\begin{equation}
    p_k = \sqrt{\frac{[s-(m_{k1} + m_{k2})^2][s-(m_{k1} - m_{k2})^2]}{4s}}.
\end{equation}

In this study, we consider a coupled-channel system reflecting the dominant physical process. The three-body process factorizes into a strong two-body interaction $D^0D^{*+} \to D^0D^{*+}$ followed by the decay $D^{*+} \to D^0 \pi^+$. Hence, the first channel represents the direct production of the observed three-body final state $D^0 D^0 \pi^+$, treated in an effective two-body formalism with asymptotic states $D^0$ and a composite $[D^0 \pi^+]$. This effective reduction is motivated by the argument that the $D^{*+}$ and $D^0\pi^+$ are physically distinct structures with different thresholds~\cite{Dai:2021wxi}. The second channel represents the intermediate resonant two-body state $D^{0}D^{*+}$. This gives the corresponding nominal thresholds $\varepsilon_1 = m_{D^0} + [m_{D^0} + m_{\pi^+}] \simeq 3869.25 \, \text{[MeV]}$ and $\varepsilon_2 = m_{D^0} + m_{D^{*+}} \simeq 3875.10 \, \text{[MeV]}$ with the masses of $D^0$, $D^{*+}$, and $\pi^+$ mesons obtained from PDG~\cite{ParticleDataGroup:2024cfk}.

Noting that our analysis focuses on the behavior of the amplitude in the vicinity of the relevant threshold $\varepsilon_k$, it is sufficient to approximate the channel momentum by the rescaled form:
\begin{equation}
    q_k =  \sqrt{s - \varepsilon_k^2}.
    \end{equation}
 
 The structure of $q_k(s)$ introduces branch points at $s =\varepsilon_k^2$. As $s$ crosses the branch cuts associated with the branch points, the sign of  $\text{Im}\,q_k(s)$ flips, giving rise to two Riemann sheets per channel. Consequently, the full two-channel amplitude is defined on a four-sheeted Riemann surface. For brevity, we adopt the Pearce and Gibson notation~\cite{PearceGibson} to label the Riemann surfaces. Specifically, we use the form $[\varsigma_1, \varsigma_2]$, where each $\varsigma_k \in \{t, b\}$ indicates the sheet [top ($t$) indicating that $\text{Im} \,q_k >0$ or bottom ($b$) implying that $\text{Im} \,q_k <0$]. Under this convention, the four Riemann sheets are labeled as $[tt]$(sheet I), $[bt]$(sheet II), $[bb]$ (sheet III), and $[tb]$ (sheet IV).
 
In this work, a primary objective is to determine the most probable pole structure for the $T_{cc}^+$ state. Probing the pole structure is essential because, in principle, the $\mathcal{S}$-matrix pole structure is a model-independent signature of a resonant state, regardless of the specific parameterization used. The physical nature of a near-threshold state is encoded in its pole structure characterized by the number of poles and the specific Riemann sheets they inhabit. This topological classification allows us to distinguish between fundamentally different physical scenarios. For instance, following the naïve pole counting criterion~\cite{Morgan:1992}, a pole in the $[bt]$ sheet generally corresponds to a molecular state while a pole-shadow pair on the $[bt]$ and $[bb]$ sheets may be associated with a compact state picture. However, tracing the pole across these sheets can be technically subtle and require careful treatment.

\subsection{Uniformization}

\begin{figure}[!ht]
    \centering
    \includegraphics[width=0.65\linewidth]{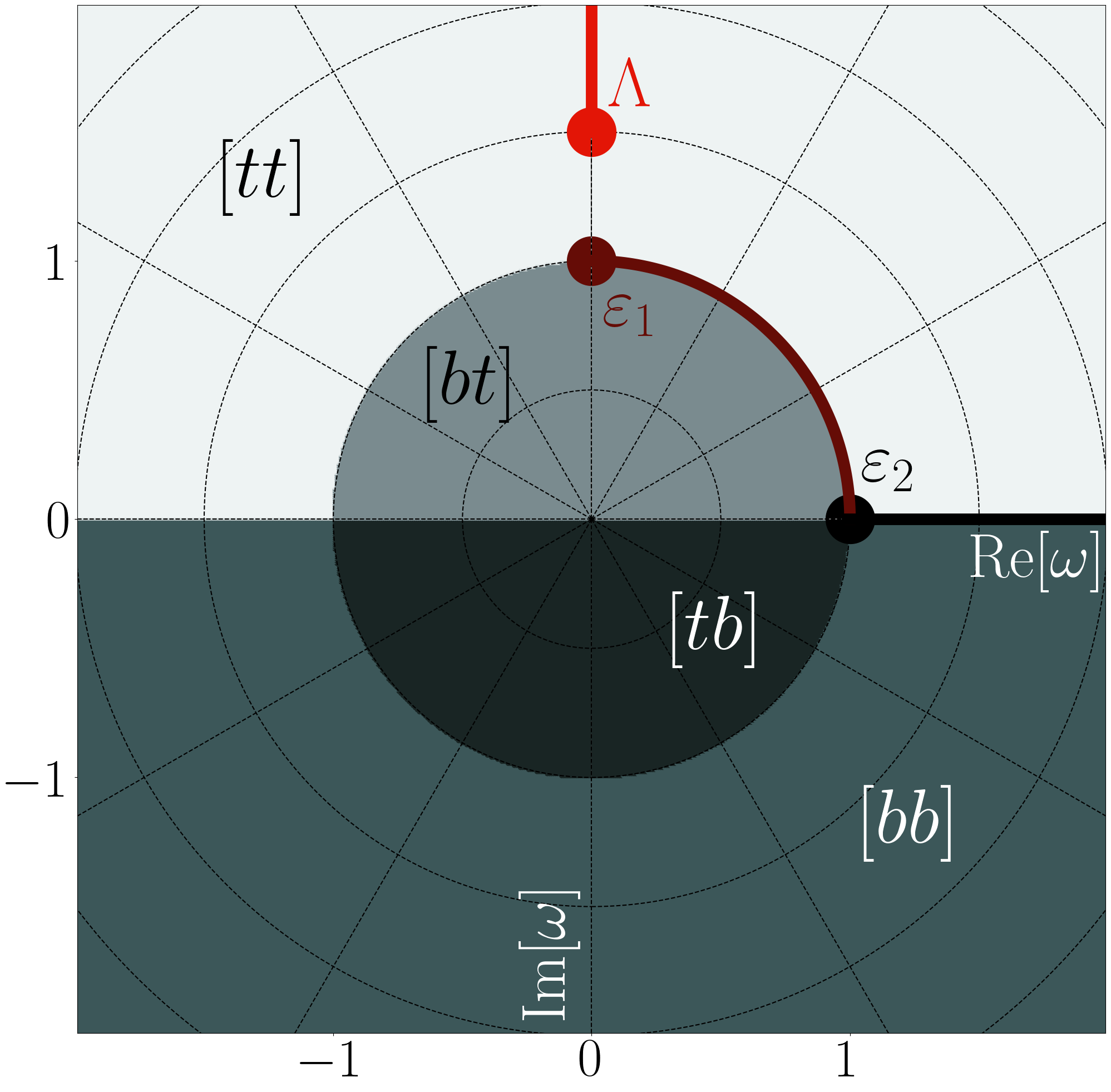}
\caption{\textbf{Mapping of the four-sheeted complex $s$-plane onto the single-sheeted complex $\omega$-plane.} The red dot labeled as $\tau$ correspond to the branch point associated with left-hand cut that arises from meson exchange. The maroon point at $\omega = i$ and the black point at $\omega = 1$ correspond to the first and second channel thresholds, $\varepsilon_1$ and $\varepsilon_2$, respectively. The physical scattering region between the thresholds is mapped onto the upper arc of the unit circle, while the region above the second threshold is represented by the segment of the real axis extending beyond $\omega = 1$.}
    \label{fig:uniformized_plane}
\end{figure}

To systematically analyze the analytic structure, we adopt a uniformizing variable $\omega$ following Kato~\cite{Kato:1965} and recent implementation in hadron spectroscopy~\cite{Santos:2023gfh}:
\begin{equation}
    \omega = \frac{q_1 + q_2}{\sqrt{\varepsilon_2^2 - \varepsilon_1^2}},
\end{equation}
which ensures that every point on the $\omega$-plane corresponds to a unique point (and sheet) in the original complex $s$-plane. In particular, the different Riemann sheets correspond to distinct angular sectors of the $\omega$-plane as shown in Fig.~\ref{fig:uniformized_plane}, allowing one to assign poles to sheets based on the location of their corresponding $\omega$-values. 

In this uniformized representation, the two-channel $\mathcal{S}$-matrix  elements take a compact and analytically controlled form:
\begin{align}
    \mathcal{S}_{11}(\omega) = \frac{\mathcal{Q}(-1/\omega)}{\mathcal{Q}(\omega)}; &\quad \mathcal{S}_{22}(\omega) = \frac{\mathcal{Q}(1/\omega)}{\mathcal{Q}(\omega)};\nonumber \\
     \det \mathcal{S} &= \frac{\mathcal{Q}(-\omega)}{\mathcal{Q}(\omega)},
\end{align}
where $\mathcal{Q}(\omega)$ is the Jost-like function that governs the pole structure of the amplitude which is explicitly given as
\begin{align}
\label{eq:Jost-like}
    \mathcal{Q}(\omega) = \prod^n \frac{1}{\omega^{2}}  &(\omega - \omega_{\text{pole}}) (\omega + \omega_{\text{pole}}^{*}) \nonumber \\
    &(\omega - \omega_{\text{reg}}) (\omega + \omega_{\text{reg}}^{*}).
\end{align}

Here, $n$ denotes the number of physical poles $\omega_{\mathrm{pole}}$ included in the model. The physical poles $\omega_{\text{pole}}$ represents a complex pole encoding a physical structure, $\omega_{\text{reg}}$ is a regulator pole that ensure correct asymptotic behavior $\mathcal{S}_{jk}\to 1$ as $\omega \to \infty$, as required by unitarity and Levinson's theorem~\cite{Newton:1982, Taylor}. The factor $\omega^{-2}$ prohibits highly energetic initial states from having no final states after the interaction. Due to hermiticity of the $\mathcal{S}$-matrix below the threshold, each pole must be paired with a symmetric pole at $-\omega_{\text{pole}}^*$. To ensure the regulator poles does not interfere with physical effects, we place them away from the scattering region, either on the $[bb]$ or $[tb]$ sheet by enforcing that $\omega_{\text{reg}} = e^{-i \pi/2}|\omega_{\text{pole}}|^{-1}$, ensuring $|\omega_{\text{pole}} \omega_{\text{reg}}| = 1$. The line shape amplitude is constructed from the $T$-matrix, which is extracted from the $\mathcal{S}$-matrix via the relation
\begin{equation}
\label{eq:ST}
    \mathcal{S}_{jk} = \delta_{jk} + 2i T_{jk},
\end{equation}
where $\delta_{jk}$ is the Kronecker delta.

\subsection{Left-Hand Cut: Phenomenological Modeling}

While the uniformized $\mathcal{S}$-matrix provides a robust framework for handling right-hand cuts and pole singularities, it does not inherently contain the dynamics of the interaction potential. In hadron physics, these dynamics are dominated by cross-channel meson exchange, which introduces a critical non-analyticity: the left-hand cut (lhc)~\cite{Eden:1966}.

The origin and location of lhcs are directly tied to the form of the potential. For a potential arising from meson exchange, which decays exponentially in coordinate space (Yukawa-type), the lhc appears along the imaginary axis in the complex momentum plane. They manifest as strips that start from $\pm i\Lambda$ where $\Lambda$ is related to the mass of the exchange particle. This was rigorously established in non-relativistic scattering theory, showing that for a potential $V(r) \sim e^{-\mu r}/r$, the scattering amplitude possesses a lhc along the imaginary $k$-axis from $k = + i\mu/2$ to $+\infty$ and a symmetric cut on the negative imaginary axis~\cite{Taylor,Newton:1982}. This structure emerges from the maximal analytic continuation of the scattering amplitude. Notably, if the inter-hadron potential were to decay faster than an exponential, no lhc would be present; however, the Yukawa-type potential of meson exchange guarantees its existence.

The phenomenological impact of the lhc is hierarchical and is cornerstone of modern hadronic amplitude analysis. The lightest meson, the pion $(\pi)$, produces the lhc closest to the physical scattering region, making it most significant for low-energy phenomenology. Exchanges of heavier mesons, such as the $\rho$ or $\omega$, generate lhcs that are situated further from the threshold and thus have a comparatively subdued effect on near-threshold phenomena. Consequently, phenomenological models must carefully incorporate these singularities to ensure the analytic structure is correctly reproduced.  This necessity is underscored by Ref.~\cite{Du:2023hlu}, which demonstrated that explicit inclusion of the lhc is pivotal for the correct determination of the nature of the $T_{cc}^+$ pole, particularly when interpreting lattice QCD data at various pion masses.

Given this imperative to include the lhc, a key challenge is to incorporate it in a manner that is both phenomenologically effective and consistent with the principles of the $\mathcal{S}$-matrix. A particularly efficient ansatz is to introduce a phase factor that directly modifies the diagonal $\mathcal{S}$-matrix elements:
\begin{equation}
\label{eq:SmatLHC}
    \mathcal{S}_{kk} \to \exp\left[ i \zeta_k \arctan \left( \frac{q_k}{\Lambda}\right)\right] \mathcal{S}_{kk},
\end{equation}
where $\Lambda$ sets the scale of the lhc effects and $\zeta_k$ parametrizes the strength of lhc contribution in each channel. The effectiveness of this approach is rooted in its direct connection to the analytic structure it aims to model. The choice of the $\texttt{arctan}$ function is mathematically deliberate; it is analytic everywhere except for branch cuts on the imaginary axis of the complex $q_k$-plane starting precisely at $\pm i \Lambda$, thereby mimicking the fundamental analyticity generated by a Yukawa potential with range of $1/\Lambda$~\cite{Taylor, Newton:1982}. \footnote{While the $\arctan$ ansatz is numerically efficient, a rigorous treatment of left-hand cut effects would require their explicit inclusion via the $N/D$ method or through unitarized dispersion relations. For this reason, throughout the remainder of this work, we refer to these contributions as simulated lhc effects.} For instance, by setting the $\Lambda \sim m_{\pi}$, the model explicitly accounts for long-range force scale associated with one pion exchange (OPE). Beyond OPE, this framework provides a generalized mechanism to capture further exchange physics, including two-pion exchange (TPE) and the contributions of heavier $\rho$ and $\omega$ mesons by adjusting the scale $\Lambda$.

In its effect on the phase shift, this ansatz introduces a smooth, energy-dependent background term, $\delta_{\text{lhc}}(q_k) = (\zeta_k/2) \arctan(q_k/\Lambda)$, which represents the cumulative contribution from the $t$-channel forces. The parameter $\zeta_k$ serves as a phenomenological strength for the left-hand cut.  A negative $\zeta_k$ , for example, yields a negative contribution to the phase shift, effectively modeling the repulsive background interactions arising from short-range $t$-channel dynamics. Crucially, this multiplicative modification is minimal and robust. It preserves the unitarity of the $\mathcal{S}$-matrix, as the factor is a pure phase and leaves analytically controlled right-hand cut structure of the original $\mathcal{S}_{kk}$ intact. It therefore serves as an efficient bridge, endowing a unitary and analytic amplitude with the necessary dynamical content from cross-channel exchanges, without the need for a complicated and computationally intensive potential model.

In summary, the central innovation of this approach lies in the parameterization of the Jost-like function $\mathcal{Q}(\omega)$ via a set of known or hypothesized pole locations $\omega_{\text{pole}}$ that can be placed freely and independently in the complex plane, allowing amplitudes to be built with arbitrary combinations in four Riemann sheets. Crucially, it enables the systemic generation of amplitudes corresponding to all physically relevant pole topologies, both with or without the inclusion of left-hand cut effects. In the next section, we leverage this theoretically controlled model space to train deep neural networks capable of inverting the scattering problem: inferring the underlying pole structure directly from the observed scattering data.

\section{Deep Learning Framework} \label{sec:3}

The model-independent construction of the $\mathcal{S}$-matrix described in the previous section enables the controlled generation of a wide variety of line shapes with explicitly assigned pole structures. Each generated amplitude corresponds to a known set of pole positions on specific Riemann sheets. This capability allows us to re-frame the inverse scattering problem as a supervised classification task: given an observable line shape, predict the corresponding pole topology class. In this section, we describe the generation of the synthetic training and validation dataset, the architecture of the deep neural network classifier, the overall classification strategy used to extract pole structures from experimental line shapes, and the inference framework to the $T_{cc}^{+}$ state.
\subsection{Dataset Construction \label{subsec:DataGen}}

\subsubsection*{Synthetic Dataset for Training and Validation}

To enable supervised classification of pole structures from experimentally accessible observables, we construct a comprehensive synthetic dataset of scattering line shapes generated within the uniformized $\mathcal{S}$-matrix formalism introduced in Section~\ref{sec:2}. Within this framework, pole singularities are introduced explicitly through the zeros of the Jost-like function (Eq.~\ref{eq:Jost-like}), allowing each generated amplitude to be labeled unambiguously by its pole content.

We define $35$ distinct pole topology classes, each characterized by a triplet $[N_{bt}, N_{bb}, N_{tb}]$, where $N_i \in [0,4]$ denotes the number of poles residing on the corresponding Riemann sheet. Notably, poles on the $[tt]$ sheet are excluded \textit{a priori} to satisfy the causality requirement, ensuring that the scattered final state cannot precede the scattering interaction~\cite{Screaton:1969he, Minerbo:1971gg} . This enlarged topology basis is motivated by the analytic structure of coupled-channel scattering amplitudes, in which poles can migrate Riemann sheets under variations of interchannel coupling and background dynamics~\cite{PearceGibson, Morgan:1992}. The resulting classes therefore span a wide spectrum of physically relevant scenarios, ranging from pole-free amplitudes to configurations involving multiple nearby poles on different sheets. By construction, this complete set ensures that no analytically allowed pole configurations is excluded \emph{a priori}, allowing the data-driven DNN analysis to discriminate among competing hypotheses without imposing model-dependent restrictions. A summary of class definitions is provided in Table~\ref{tab:pole_class}.

Crucially, to isolate the spectroscopic impact of the left-hand cut, we generated two complementary ensembles of scattering amplitudes: a baseline set of scattering amplitudes without lhc effects, and a set of scattering amplitudes incorporating the lhc effects via the ansatz in Eq.~\eqref{eq:SmatLHC}. We normally sampled the cut scale $\Lambda = 280 \pm 50$ [MeV], a range physically motivated by the need to capture TPE as the dominant long-range interaction in the $s$-wave $DD^*$ sector. This scale aligns with the pion masses used in lattice QCD studies, allowing for a more consistent comparison between our analysis and first-principles calculations. Meanwhile, the channel-specific strength $\zeta_k \in [-1, 1]$ was sampled uniformly to probe the wide spectrum from repulsive to attractive background interactions.

Synthetic amplitudes are constructed over the energy window $[3871, 3900]$~MeV, aligned with the kinematic region probed by LHCb measurements of the $D^0 D^0 \pi^{+}$ final state~\cite{LHCb:2021vvq, LHCb:2021auc}. This interval is discretized into $57$ uniform bins, matching the experimental resolution. For each amplitude, we sample a single energy value per bin using a uniform distribution, simulating the finite resolution inherent in experimental detection. This bin-wise randomization emulates detector-level smearing and introduces variability across the dataset.

The resulting observable line shapes are computed from the differential event rate:
\begin{equation}
\label{eq:dN/ds}
    \frac{\mathrm{d}N}{\mathrm{d}\sqrt{s}} = \rho(\sqrt{s})\left[ \left|T(\sqrt{s}) \right|^2 + b(\sqrt{s}) \right],
\end{equation}
where $\rho(\sqrt{s})$ denotes the two-body phase space factor and $b(\sqrt{s})$ models non-resonant background contributions. The phase space factor $\rho(\sqrt{s})$ is taken as a kinematically consistent function based on the Källén triangle function:
\begin{equation}
\label{eq:rho}
    \rho(\sqrt{s}) = \frac{\lambda^{1/2}(s, m_{D^0\pi^+}^2, m_{D^0}^2)}{2\sqrt{s}}.
\end{equation}

To enhance synthetic dataset diversity, the background term $b(\sqrt{s})$ is randomly chosen from predefined functional forms: (i) a quadratic polynomial, (ii) an exponential decay, or (iii) zero contribution. This stochastic background selection introduces nontrivial shape variations and challenges neural networks to focus on pole-driven features rather than background-specific patterns.

The full scattering amplitude $T(\sqrt{s})$ entering Eq.~\eqref{eq:dN/ds} is given by
\begin{equation}
    T(\sqrt{s}) = \sum_{k=1}^{2} \alpha_k \,T_{1k}(\sqrt{s}),
\end{equation}
where $\alpha_k$ denotes the production coefficients modeling the relative rates for the elastic and inelastic channels. The inclusion of the off-diagonal term $T_{12}$, which encodes interchannel coupling, is crucial for capturing interference patterns and resolving pole configurations that may be degenerate in the diagonal elements alone~\cite{Santos:2023gfh}.

Each pole is generated as a zero of a Jost-like function defined in Eq.~\eqref{eq:Jost-like} and is constrained to reside near the $D^0D^{*+}$ threshold. Specifically, pole positions $\sqrt{s}_{\text{pole}}$ are sampled within the following bounds:
\begin{equation}
    \begin{cases}
        \varepsilon_2 - 5.0 \leq \, \text{Re}\, \sqrt{s}_{\text{pole}} \, \, \leq \varepsilon_2 + 10.0,\\
        \quad \quad \,0.0 \leq |\text{Im}\, \sqrt{s}_{\text{pole}}| \leq 20.0.
    \end{cases}
\end{equation}

This ensures the poles are located within the physically relevant region influenced by threshold effects and coupled-channel dynamics.

Specifically, we construct a $114$-dimensional input vector by concatenating two components: the discretized COM energy values $\{ \sqrt{s_i}\}$ and their corresponding differential event rates $\{ \mathrm{d}N/\mathrm{d}\sqrt{s_i}\}$ across $57$ bins. This dual encoding enhances the expressivity of the input representation, enabling the network to capture subtle variations across the line shape. The complete dataset is composed of labeled pairs $(x^j, y^j)$, where $y^j \in \{n \in \mathbb{Z} \, | \, 0 \le n \le 34\}$ denotes the class label corresponding to one of the $35$ pole configurations, resulting in a dataset of $350{,}000$ line shapes. For each class, we generate $10^4$ synthetic samples, resulting in a well-balanced dataset suitable for supervised learning. The data is then randomly split into training and validation subsets using a $4{:}1$ ratio, enabling consistent training and robust out-of-sample performance evaluation.
\begin{figure*}[!ht]
    \centering
    \includegraphics[width=0.75\linewidth]{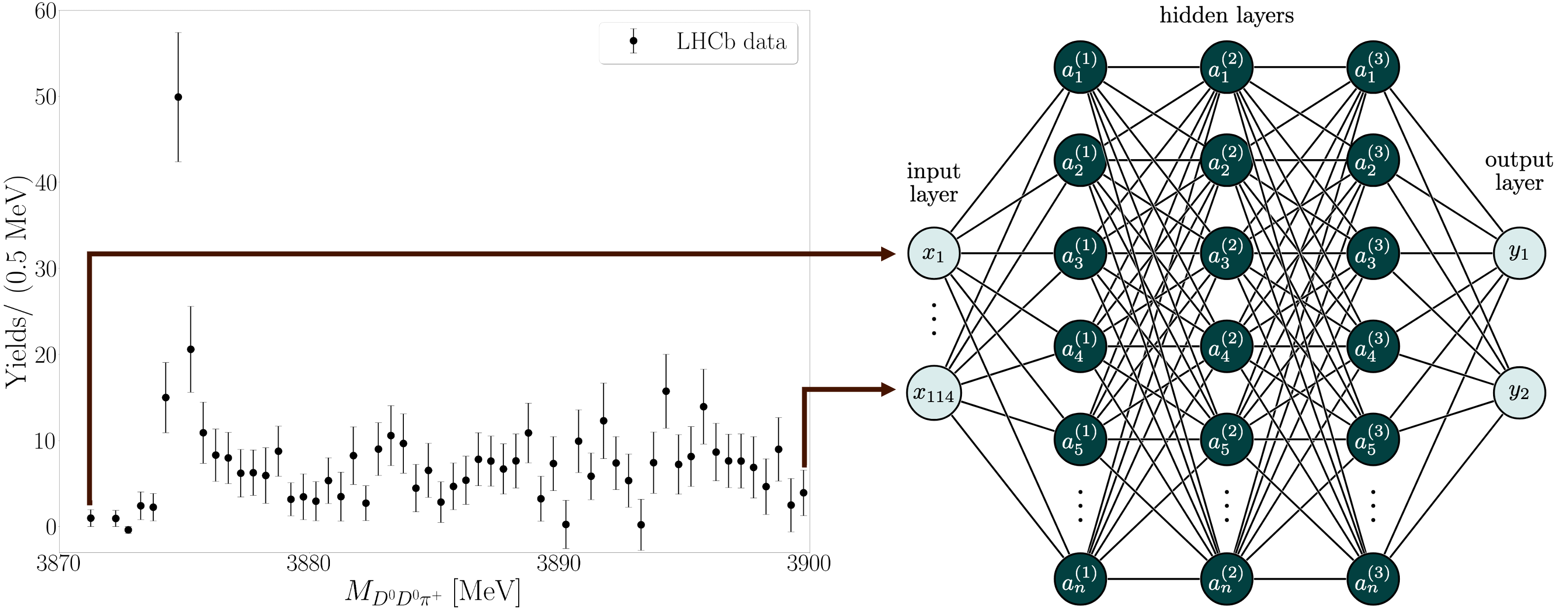}
    \caption{\textbf{Schematic workflow for inference using trained DNN classifiers on empirical data.} (Left) The experimental invariant mass distribution for the $D^{0}D^{0}\pi^{+}$ final state as reported by the LHCb Collaboration\cite{LHCb:2021vvq,LHCb:2021auc}, including statistical and detector-induced uncertainties. (Right) A diagram of trained DNN model with an input layer of $114$ nodes corresponding to concatenated energy-amplitude pairs. The illustrated model contains three hidden layers, though architectures with up to five hidden layers were also explored. The output layer contains two nodes, the dimensionality was reduced from $35$ as a result of preliminary coarse-grained classification and finalist selection strategy. The LHCb data was extracted from~\cite{hepdata.Tcc}, and the TikZ code for the DNN figure was taken from~\cite{neutelings:2021}; both are reproduced under the terms of the Creative Commons Attribution $4.0$ International License (CC BY $4.0$).}
    \label{fig:empirical_inference}
\end{figure*}

\subsubsection*{Empirical Dataset for Inference}

To apply the trained DNN classifiers to the real-world data, we extract the experimental line shape from the $D^{0}D^{0}\pi^{+}$ invariant mass distribution reported by the LHCb Collaboration\cite{LHCb:2021auc, LHCb:2021vvq}. Each spectrum point includes a central energy with uncertainties on the $x$-axis and a central amplitude with uncertainties on the $y$-axis. These uncertainties arise from both statistical fluctuations and detector resolution effects. To address this, we construct inference-ready samples that reflect the statistical and systematic uncertainties of the LHCb spectrum, as shown in the left panel of Figure~\ref{fig:empirical_inference}.

To emulate realistic detector-level effects, we generate independent realizations of the experimental spectrum. For each pseudo-empirical realization, one energy-amplitude pair is sampled per reported data point. Energy values are uniformly sampled from the per-event uncertainty range. To simulate amplitude fluctuations, we explore two uncertainty models. First, the amplitude values are drawn from a Gaussian distribution centered at the reported value with a standard deviation equal to one-third of the quoted upper uncertainty. To complement this, we also implement a uniform sampling strategy over the full error band of the amplitude values. This assumption is more conservative and agnostic, avoiding any bias to the central value.

For each realization, the corresponding differential event rates $\mathrm{d}N/\mathrm{d}\sqrt{s}$ are taken directly from the experimental bin counts with no further smoothing or rescaling. The sampled energy values and bin heights are then concatenated to form an input vector $x^{\text{exp}} \in \mathbb{R}^{114}$, consistent with the input format used during training. Crucially, these sampling procedures are carried out independently for each trained DNN. That is, each DNN receives its own unique set of $1{,}000$ empirically-derived line shapes, ensuring that the final class probabilities reflect both model uncertainty from training variations across DNNs and data uncertainty from fluctuations in the empirical spectrum.

With both the synthetic datasets for supervised learning and the protocol for LHCb-based inference inputs outlined, we now proceed to the design of the deep neural network architecture.

\subsection{Neural Network Architecture and Design}

To map synthetic line shapes into their corresponding pole configuration classes, we employ supervised learning using deep neural networks (DNNs). Each input vector $x^j \in \mathbb{R}^{114}$ is fed into a fully connected feed-forward network $f_\theta$, parameterized by a set of weights and biases $\theta$. The network outputs a probability vector over $\mathcal{Z}$ number of classes via a softmax activation function~\cite{bridle1989, bishop2006prml, Goodfellow-et-al-2016}:
\begin{equation}
    \hat{y}^j = \text{Softmax}[f_\theta(x^j)] \in \mathbb{R}^{\mathcal{Z}},
\end{equation}
with the predicted class given by $\text{arg max}_l\, \hat{y}_l^j$. Training the models are performed by minimizing the categorical cross-entropy loss  function~\cite{Goodfellow-et-al-2016, Murphy2012ML}, defined as:
\begin{equation}
    \mathcal{L} = - \sum_{l=1}^{\mathcal{Z}} y_l \log (\hat{y}_l),
\end{equation}
where $y_l \in \{0,1\}$ is the true one-hot encoded true class label and $\hat{y}_l \in [0, 1]$ is the predicted class probability. To improve convergence and expressiveness, each hidden layer employs the Rectified Linear Unit (ReLU) activation function~\cite{Nair2010ReLU} :
\begin{equation}
    \text{ReLU}(z) = \text{max}(0, z),
\end{equation}
which has been shown to promote sparsity and mitigate the vanishing gradient problem~\cite{Glorot:2011}.

To ensure robust and interpretable classification of pole configurations, we implemented two complementary approaches to neural network design and tuning: (1) a manual grid search, where model architectures and hyperparameters were explicitly specified, and (2) an automated, adaptive search using the Optuna optimization framework~\cite{optuna_2019}. These approaches enabled a broad and deep exploration of the hyperparameter and architecture space, providing a diverse ensemble of models for downstream inference.

All models were implemented using PyTorch~\cite{Paszke:2019xhz} and evaluated on a held-out validation set using standard classification metrics. Each model was trained using the Adam optimizer~\cite{Kingma2015Adam} with appropriate regularization imposed via dropout~\cite{Srivastava2014Dropout} and weight decay~\cite{Krogh:1991}.

In manual grid search strategy, each model was selected by specifying the number of hidden layers $h_1 \in [2,5]$, neurons per layer $h_2 \in \{100x_0 \, | \, x_0=2, 3, \cdots, 8\}$, dropout $h_3 \in [0.025, 0.200]$. The learning rate was chosen as $h_4 = 10^{-5}$ and the weight decay $h_5 = 10^{-5}$ all throughout. This grid search allowed us to establish baseline architectures that are both interpretable and tunable. 

To complement the manual design process and explore a broader space of architectures and hyperparameters, we employed Optuna~\cite{optuna_2019}, a modern hyperparameter optimization framework based on Bayesian sampling and pruning strategies~\cite{snoek2012practical, Bergstra:2011}. For each trial, Optuna searched the architecture and training parameter space: $h_1 \in [3,5]$, sampled uniformly $h_2 \in [200, 600]$, sampled from the continuous range $h_3 \in [0.0, 0.3]$, sampled log-uniformly $h_4 \in [10^{-5}, 10^{-3}]$, and $h_5 \in [10^{-6}, 10^{-3}]$. The activation function was fixed to ReLU~\cite{Nair2010ReLU} for compatibility and comparability. Each Optuna trial involved training a candidate architecture for $75$ epochs on the training set. The Median Pruner was enabled to automatically terminate under-performing trials early. This pruning strategy saved compute time and focused optimization on promising regions of architecture space. Importantly, Optuna trials are required to achieve $\geq 45\%$ validation accuracy within the $75$-epoch window to be retained for downstream ensemble use. 

This two-pronged strategy, manual tuning for interpretability and Optuna for adaptive exploration, yielded a robust ensemble of $216$ trained DNN models. Their combined insights enabled a deeper understanding of the pole structure of the $T_{cc}^{+}$ state and formed the basis for the experimental inference. All codes used in this analysis are openly available for reproducibility in~\cite{Tcc_ML}.

\begin{figure}[!ht]
    \centering
    \includegraphics[width=0.98\linewidth]{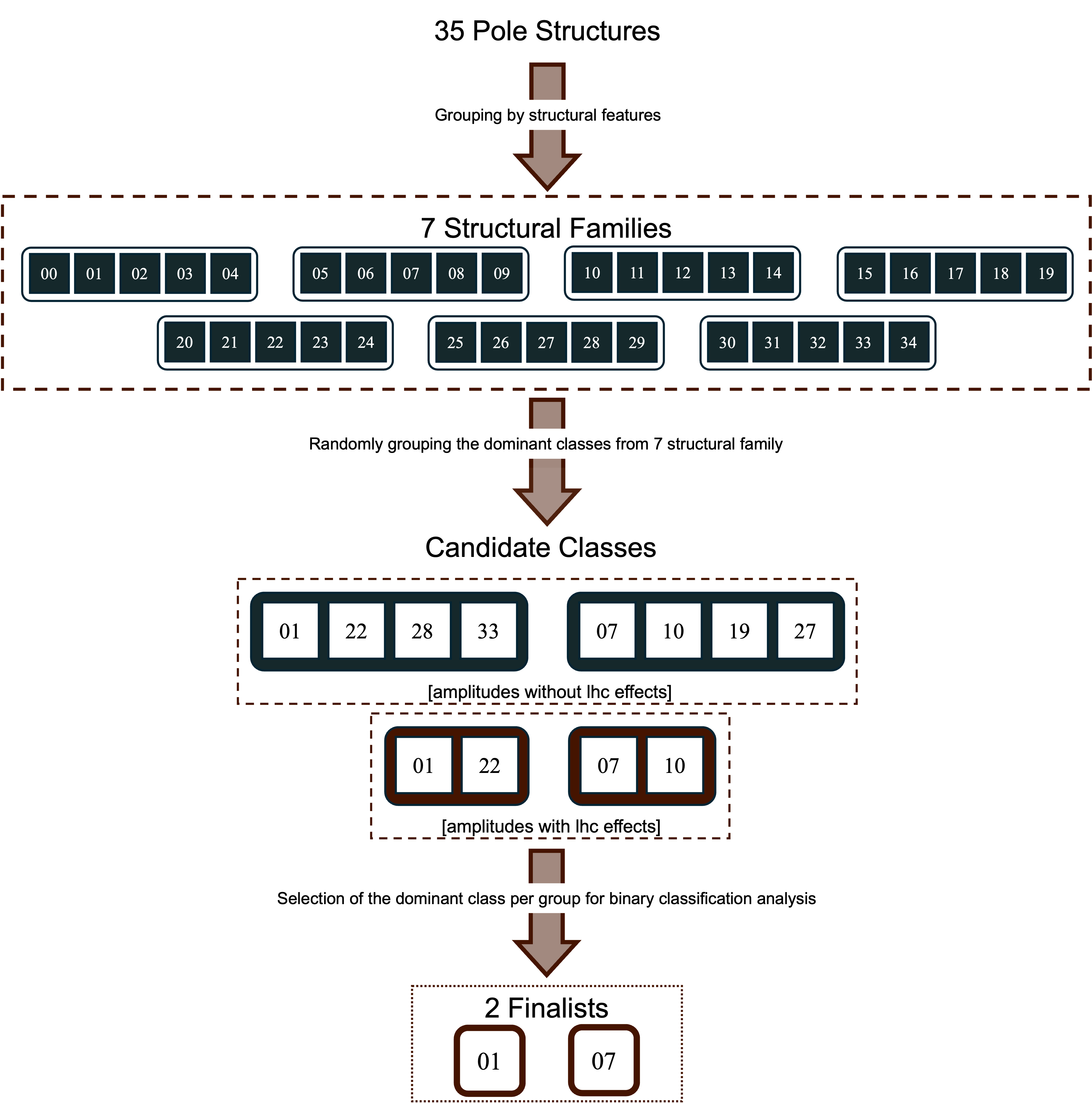}
    \caption{\textbf{Schematic overview of the coarse-grained classification and finalist selection strategy.} The original $35$ pole classes (top row) were grouped into seven structural families (second row) to address degeneracies in the line shapes. Dominant classes from each family (third row) formed an $8$-class candidate set for amplitudes without lhc effects and $4$-class candidate set for amplitudes with lhc effects. These classes were randomly grouped into two sets to select the dominant class from each set (last row) for an extensive binary classification analysis using an ensemble of $50$ DNNs.}
    \label{fig:finalistselection}
\end{figure}

\subsection{Coarse-Grained Classification and Finalist Class Selection}

\quad Our initial classification task involved distinguishing among $35$ distinct pole configurations, each representing a unique arrangement of singularities across the $[bt]$, $[bb]$, and $[tb]$ Riemann sheets. However, the attempts to train $3$ DNN classifiers with amplitudes (sans lhc effects) directly on this $35$-output node architecture showed limited success. In particular, the training accuracy plateaued below $60\%$ and validation performance stagnated. This indicated that the classification scheme was too fine-grained relative to the degeneracies present in the line shapes. Specifically, several pole configurations led to morphologically indistinguishable patterns and consequently confounded the DNN model's ability to learn meaningful distinctions. Moreover, the computational burden of training DNNs on such large output space proved to be inefficient. Hence, the DNN models are expected to be incapable of  reliably inferring the pole structure of the $T_{cc}^{+}$ from the LHCb data.

To address this, we implemented a coarse-graining strategy (Figure~\ref{fig:finalistselection}) that grouped the $35$ pole configurations into seven broader structural families based on shared topological features including the number and type of poles on each Riemann sheet. This structural grouping was organized as follows: $\text{SF}1$: classes $00-04$, corresponding to the simplest pole topologies, $\text{SF}2$: classes $05-09$, involving up to two poles distributed across different sheets. Moreover, $\text{SF}3$: classes $10-14$ and $\text{SF}4$: classes $15-19$, features three-pole configurations with different sheet distributions. Finally, $\text{SF}5$: classes $20-24$, $\text{SF}6$: classes $25-29$, and $\text{SF}7$: classes $30-34$ contains the most complex arrangements with up to four poles. One may refer to Table~\ref{tab:pole_class} for a summary of this structural categories.

\begin{figure*}[!ht]
    \centering
    
    \subfloat[\label{fig:SFaccuracy1}]{
        \includegraphics[width=0.47\linewidth]{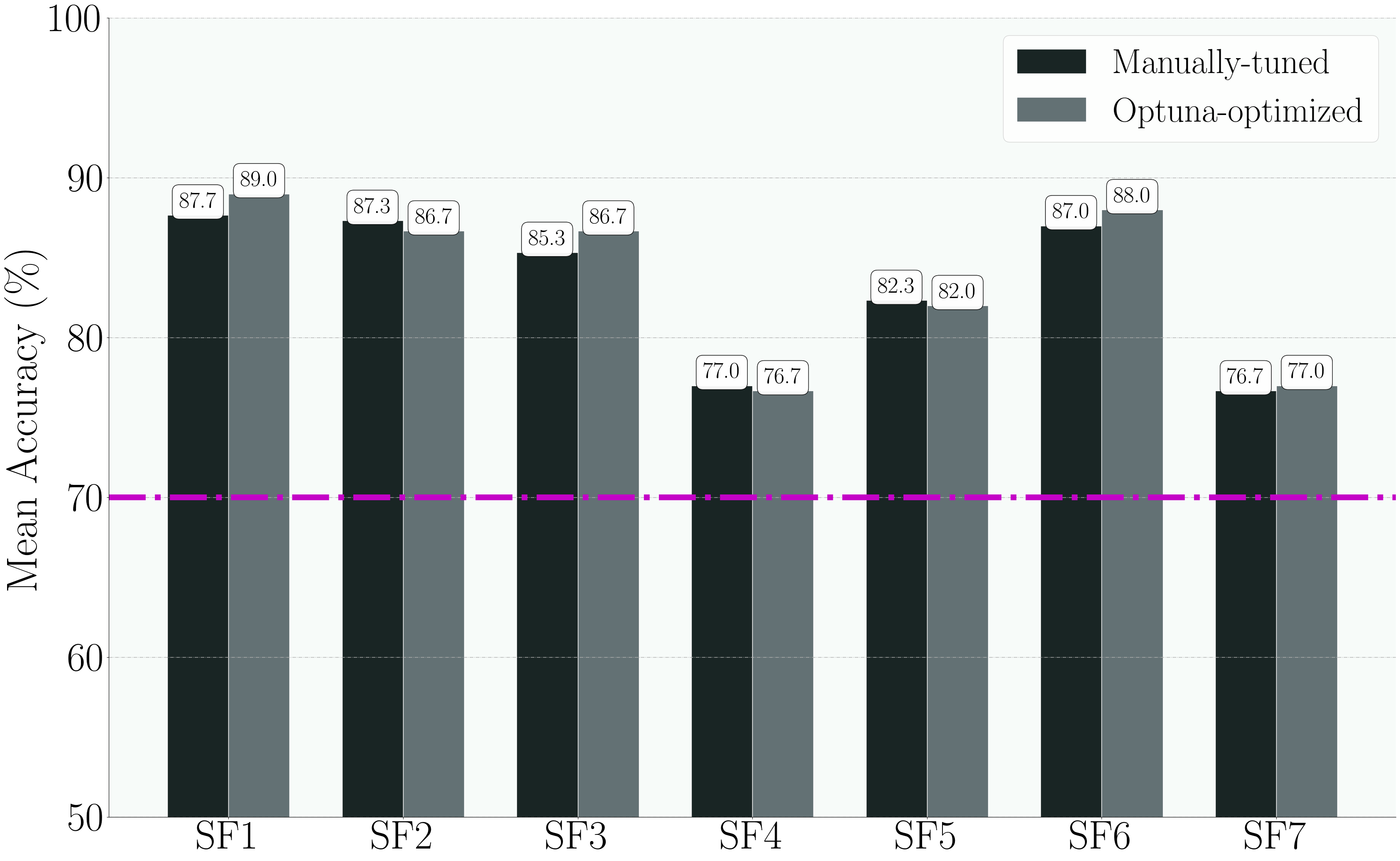}
    }
    \hfill
    \subfloat[\label{fig:SFaccuracy2}]{
        \includegraphics[width=0.47\linewidth]{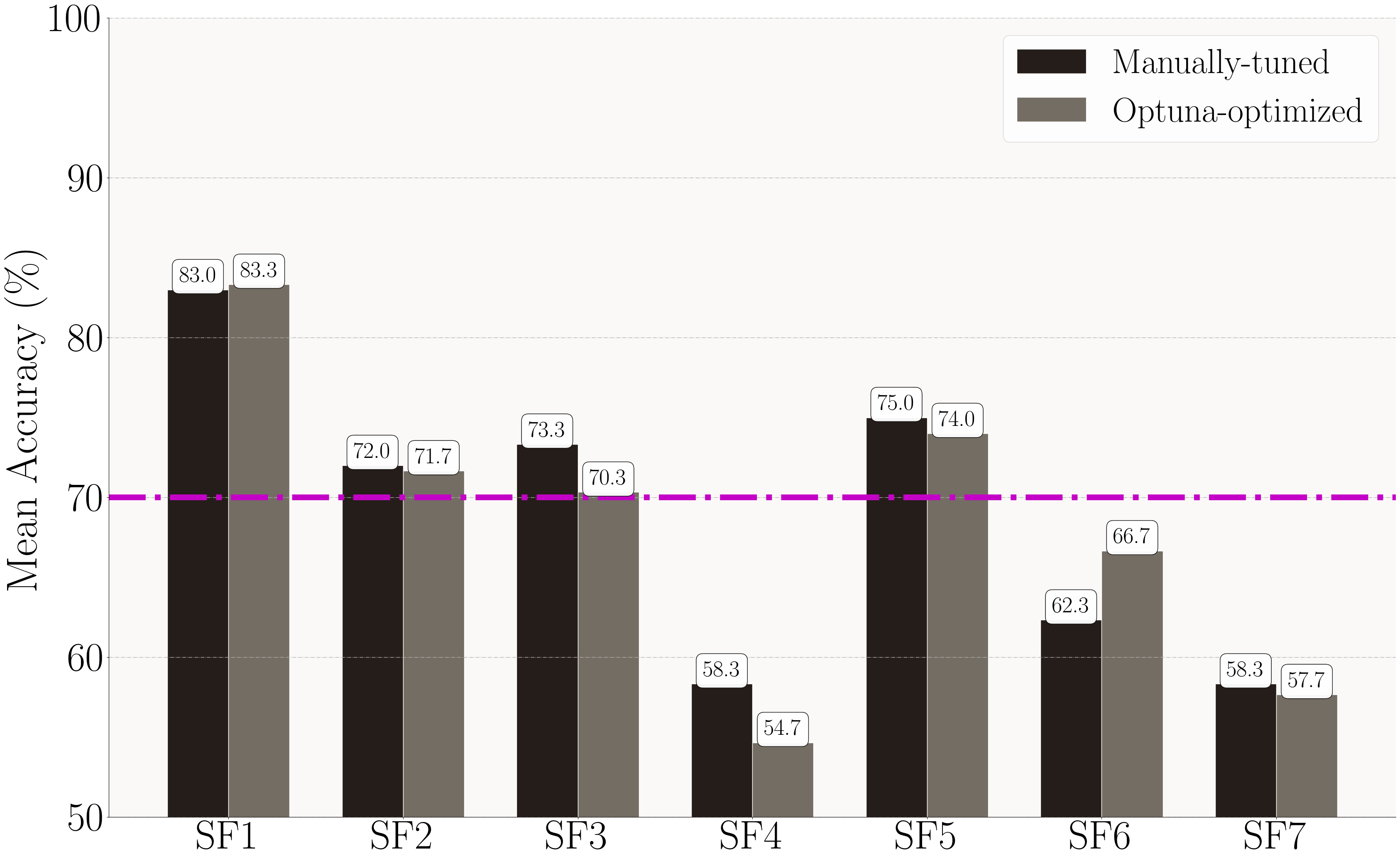}
    }
    
    \caption{\textbf{Mean validation accuracy of DNN models across $7$ structural families.} Each bar represents the average over three independently-trained manually tuned and Optuna-optimized DNN models per family. The left panel shows the classification performance of DNNs trained on amplitudes without lhc effects and the right panel displays the classification performance of DNNs trained on amplitudes with lhc effects. The fuchsia horizontal dashed-dotted line marks the $70\%$ validation accuracy threshold; \textit{i.e.}, predictions below this threshold are insufficiently reliable for subsequent inference using LHCb data.}
    \label{fig:SFaccuracy}
\end{figure*}

For the amplitudes without lhc effects, we trained six independent DNNs ($3$ are manually-tuned and the other $3$ are Optuna-optimized) for each of the seven structural families on a $5$-output node classification task. To explore the effects of model complexity, we employed architectures with increasing depth and width.  This ensemble strategy, from simple to more expressive networks, enables us to robustly generalize the results and mitigate potential overfitting associated with any single architecture. Building upon this, Fig.~\ref{fig:SFaccuracy1} compares the mean validation accuracy achieved by the manually-tuned and Optuna-optimized DNN models for each of the seven structural families. The results show that hyperparameter and architecture optimization via Optuna yields modest but consistent improvement in accuracy for most families, particularly in $\text{SF1}$, $\text{SF3}$, and $\text{SF6}$, where gains of about $1{-}2\%$ are evident. These families also exhibit the highest overall accuracies, reflecting a clearer separation between candidate configurations within these structural categories. In contrast, $\text{SF2}$ and $\text{SF5}$ show minimal differences between manual and Optuna tuning, suggesting that model performance in these families is less sensitive to hyperparameter adjustments. The lowest accuracies occur for $\text{SF4}$ and $\text{SF7}$, regardless of DNN design strategy, due to pronounced ambiguities between specific pole configurations. In $\text{SF4}$, the primary confusion arises between $[0_{bt}, 2_{bb}, 1_{tb}]$ and $[0_{bt}, 1_{bb}, 2_{tb}]$ pole topologies, as well as between $[1_{bt}, 0_{bb}, 2_{tb}]$ and $[1_{bt}, 1_{bb}, 1_{tb}]$, which hampers the DNN classifier's ability to confidently separate these cases. Similarly, in $\text{SF7}$, misclassifications are largely driven by overlapping features between $[2_{bt}, 0_{bb}, 2_{tb}]$ and $[2_{bt}, 1_{bb}, 1_{tb}]$, as well as between $[1_{bt}, 2_{bb}, 1_{tb}]$ and $[1_{bt}, 1_{bb}, 2_{tb}]$ pole topologies. These structural families thus represent inherently more challenging classification tasks, where subtle differences in the line shapes reduces DNNs discriminative power.

It is worth noting however, that despite these average gains, many of the individual Optuna-optimized DNN models we tested were not ultimately helpful because they tended to overfit. In our Optuna setup, candidate architectures are selected based on achieving high validation accuracy at relatively early epochs, a criterion that can favor models which quickly memorize the training data but fail to generalize when trained to convergence. This reflects a well-documented issue in neural architecture search (NAS), where reliance on low-fidelity evaluations or short-horizon validation signals often misranks candidate architectures~\cite{Elsken:2019}. More broadly, this challenge reflects the inherent tension between model expressivity and generalization: highly expressive architectures discovered through automated search can easily overfit unless balanced by appropriate regularization or multi-objective selection criteria. Nonetheless, the searched models that failed to generalize were systematically rejected from our analysis.

When the amplitude samples from the LHCb spectrum is fed to the DNNs trained on amplitudes without lhc effects, this strategy yielded eight dominant classes. In particular, class~$01$ ([$1_{bt}, 0_{bb}, 0_{tb}$]) consistently dominated in the first structural family $\text{SF}1$. Meanwhile, class~$07$ ([$1_{bt}, 1_{bb}, 0_{tb}$]) was strongly favored in $\text{SF}2$. In $\text{SF}3$ and $\text{SF}4$, the leading configurations were class~$10$ ([$3_{bt}, 0_{bb}, 0_{tb}$]) and class~$19$ ([$1_{bt}, 1_{bb}, 1_{tb}$]), respectively. The pole configuration [$1_{bt}, 1_{bb}, 1_{tb}$] had previously been proposed by~\cite{Santos:2024bqr} as a leading pole structure for the $P_{\psi}^{N}(4312)^{+}$, hinting at a recurring structural pattern in near-threshold exotic states. From $\text{SF}5$, class~$22$ ([$0_{bt}, 0_{bb}, 4_{tb}$]) was identified, suggesting that in some cases, multiple poles on a single sheet can approximate the amplitude structure. In $\text{SF}6$, despite achieving a relatively high training and validation accuracy of approximately $87\%$ across $6$ DNN models, the DNNs could not decisively select between class~$27$ ([$1_{bt}, 0_{bb}, 3_{tb}$]) and class~$28$ ([$0_{bt}, 1_{bb}, 3_{tb}$]). This ambiguity likely stems from a structural mismatch between the classes in $\text{SF}6$ and the empirical features of the LHCb data, compelling the DNNs to distribute their predictions across suboptimal class candidates. Nevertheless, both classes were retained in the subsequent stage to ensure that no potentially viable configuration was prematurely ruled out. Finally, in $\text{SF}7$, class~$33$ ([$1_{bt}, 2_{bb}, 1_{tb}$]) was favored, a configuration also suggested in~\cite{Frohnert:2025usi} for the $P_{\psi}^{N}(4312)^{+}$ state, further strengthening the hypothesis of structural analogies among exotic hadrons.

When the same analysis is repeated using amplitudes that include lhc effects, the overall classification performance follows the same qualitative trends but with significantly reduced accuracies across all structural families. This is clearly visible in Fig.~\ref{fig:SFaccuracy2}, where the mean validation accuracies fall by approximately $5{-}20\%$. The decrease is most pronounced in $\text{SF4}$, $\text{SF6}$, and $\text{SF7}$, whose accuracies drop below $70\%$ regardless of whether the models are manually-tuned or Optuna-optimized. Even families that perform comparatively well in the case where lhc effects are not incorporated, such as $\text{SF3}$ and $\text{SF6}$, exhibit a substantial loss when lhc effects are present, reflecting the increased overlap in the line shapes. In contrast, $\text{SF1}$ and $\text{SF5}$ remain the most robust even under lhc contamination, with accuracies that stay marginally above $70\%$ accuracy. Consequently, the choice to set the $70\%$ reliability threshold was made after a thorough inspection of the confusion matrices from the trained classifiers, and thereby rejecting the pole topology classes involved in these structural families. Thus, for the DNN analysis involving amplitudes with lhc effects, only the pole topologies $[1_{bt}, 0_{bb}, 0_{tb}]$, $[1_{bt}, 1_{bb}, 0_{tb}]$, $[3_{bt}, 0_{bb}, 0_{tb}]$, and $[0_{bt}, 0_{bb}, 4_{tb}]$ was considered for the subsequent class-selection stage.

To further refine the analysis, we randomly partitioned the dominant candidate configurations into two groups for comparative evaluation. In particular, for the case involving amplitudes without lhc effects, we have $\text{G}1: \text{classes}\,\{01, 22, 28, 33\}$ and $\text{G}2: \text{classes}\,\{07, 10, 19, 27\}$.  Meanwhile, for the dominant candidate classes associated with amplitudes contaminated by lhc effects, we performed two separate binary classification analysis such that $\text{G1: classes \{01, 22\}}$ and $\text{G}2: \text{classes}\,\{07, 10\}$ . Each group was treated as a standalone classification task, where the objective was to identify the most likely configuration within the group that best explains the LHCb data. For each group, we trained $8$ independent DNN models ($4$ are manually-tuned and the remaining $4$ models are Optuna-optimized), each operating on a four-output and binary-output node architecture corresponding to the configurations within its assigned set. 

Upon inference using the LHCb data, the classifiers trained both on amplitudes with and without lhc effects consistently identified class~$01$ ([$1_{bt}, 0_{bb}, 0_{tb}$]) as the most probable configuration within $\text{G}1$, and class~$07$ ([$1_{bt}, 1_{bb}, 0_{tb}$]) as dominant in $\text{G}2$. These two configurations thus advanced as the finalist candidates, representing two structurally distinct hypotheses for the nature of the $T_{cc}^+$ state.

\subsection{Binary Classification Analysis: Molecular vs. Compact Interpretations}

\begin{figure}[!ht]
    \centering
    
    \subfloat[\label{fig:binaryPCD1}]{
        \includegraphics[width=0.47\linewidth]{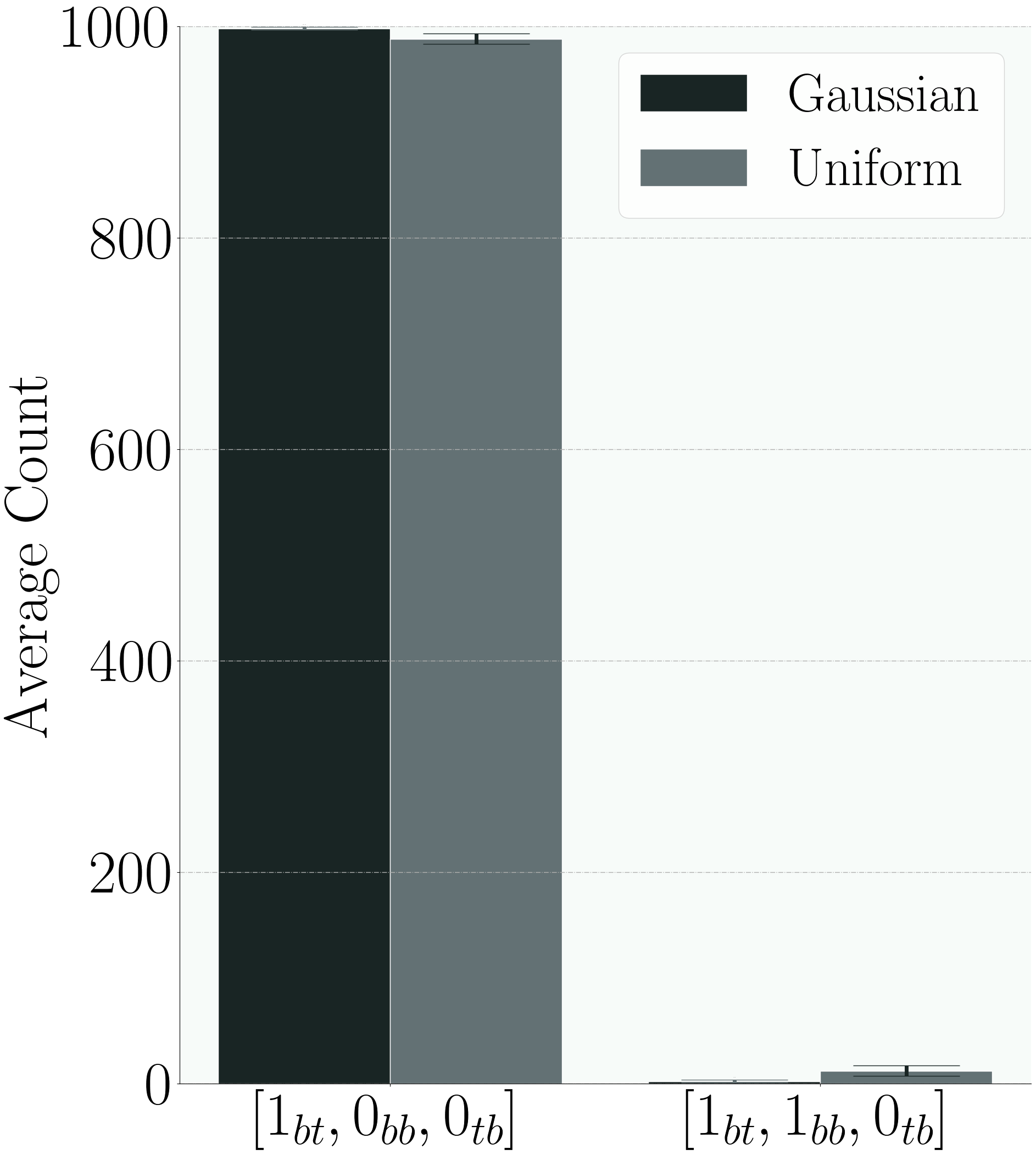}
    }
    \hfill
    \subfloat[\label{fig:binaryPCD2}]{
        \includegraphics[width=0.47\linewidth]{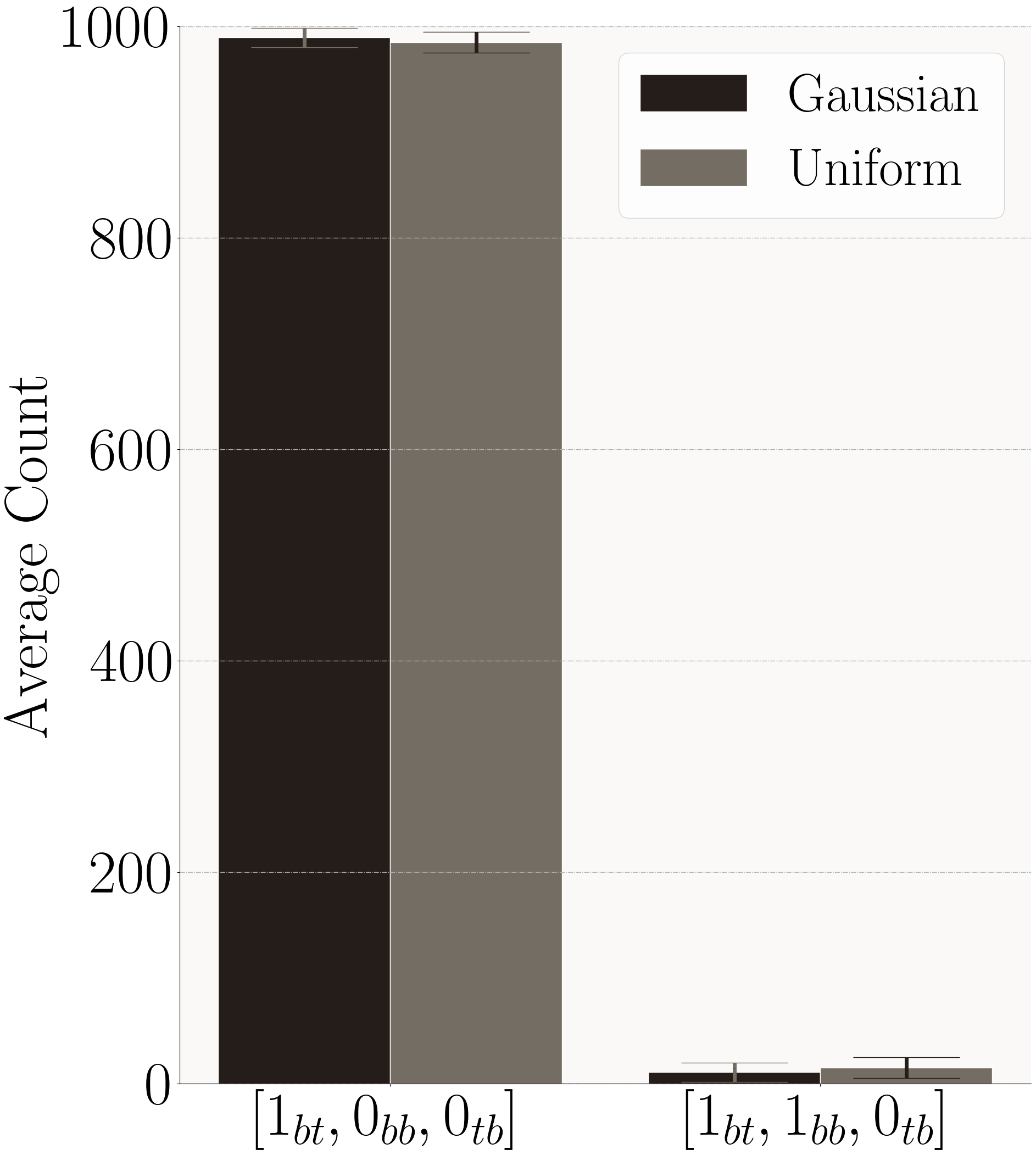}
    }
    
    \caption{\textbf{Molecular vs. compact tetraquark picture.} Ensemble-averaged predicted class distributions from $100$ DNN models trained on (a) amplitudes without lhc effects and (b) on amplitudes with lhc effects. Predictions are made on LHCb data using two amplitude sampling assumptions: Gaussian and uniform distribution. The mean classification counts are obtained from $50{,}000$ LHCb amplitude realizations, with predictions aggregated across an ensemble of $50$ DNNs. Error bars represent standard errors of the ensemble means.}
    \label{fig:binaryPCD}
\end{figure}

In the final stage of the classification pipeline, we conducted an extensive analysis comparing the two finalist configurations: class~$01$ ([$1_{bt}, 0_{bb}, 0_{tb}$]) and class~$07$ ([$1_{bt}, 1_{bb}, 0_{tb}$]). To ensure robust and statistically significant results, we trained an ensemble of $50$ independently initialized DNN models, each tasked with a binary classification between the two pole topologies. Consistent with the earlier subgroup training results, where manual tuning performed on par with Optuna-optimized models, we adopted only the manual tuning strategy at this stage. In the binary classification setting, the Optuna optimization framework loses its practical advantage since the DNN models trained on amplitudes (sans lhc effects) typically converge to near-maximum validation performance after $\sim 300-500$ epochs, rendering the long search cycles required by Optuna computationally prohibitive. The manually tuned ensemble was deliberately diversified to explore the model space thoroughly, with architectures comprising $2{-}5$ hidden layers and the number of neurons per hidden layer are selected in the range $[200,900]$ (see Table~\ref{tab:DNN_summary} for the complete list of the DNN models). Across this ensemble, validation accuracies hovered in the $93-95\%$ range for DNNs trained on amplitudes without lhc effects and $81-87\%$ classification accuracy for DNNs trained on amplitudes with lhc effects, underscoring both the robustness of the manual tuning approach and the high separability of the two pole topologies in the reduced hypothesis space.  

The final ensemble-averaged class distribution with the corresponding standard error of the mean are shown in Figure~\ref{fig:binaryPCD}. Under Gaussian and uniform sampling strategies in the intensity axis of the LHCb data, the pole configuration $[1_{bt}, 0_{bb}, 0_{tb}]$ is overwhelmingly favored, the class which correspond to a single pole on the $[bt]$ sheet. These results exhibit both sharp dominance and low ensemble variance, indicating strong agreement across models and sampling assumptions. Notably, in contrast to conventional fitting procedures, such as the $\Lambda(1405)$ analysis in~\cite{Yamada:2021cjo}, where increasing model complexity (e.g., by introducing more poles and free parameters) naturally improves the fit quality as measured by the reduced chi-square $\chi_\nu^2$, our trained DNNs consistently favor the simplest adequate explanation of the LHCb data. It avoids overfitting by rejecting more complex pole topologies that might artificially improve agreement with fine features of the line shape. For instance, the pole configuration $[1_{bt}, 1_{bb}, 0_{tb}]$ which includes additional pole on the $[bb]$ sheet, could in principle refine the line shape and absorb residual binning or resolution effects. However, this more ambiguous and flexible topology is not selected by the classifiers, demonstrating the model's preference for parsimony and its insensitivity to artifacts that might otherwise bias a traditional fit. 

This minimal inference strengthens the robustness of the result and supports a topologically grounded interpretation of the $T_{cc}^{+}$ as a hadronic molecule. However, the interpretation that the $T_{cc}^+$ is a loosely bound state requires careful consideration at this point, since a virtual state pole of the $D^0D^{*+}$ channel may traverse onto the $[bt]$ sheet due to strong interchannel coupling~\cite{Frazer:1964, PearceGibson, Fernandez-Ramirez:2019koa}. Despite this distinction, bound and virtual states are encompassed within the broader classification of hadronic molecules~\cite{Matuschek:2020gqe}. In any case, a reliable interpretation must be substantiated by the corresponding pole trajectory in the complex plane, which can only be established using model-dependent methods~\cite{Hanhart:2014ssa, Baru:2003qq, Badalyan:1982}.

To close this section, while the classification results unambiguously point to an isolated pole in the $[bt]$ sheet, the neural network framework is inherently agnostic to underlying dynamics that generate such a pole.  In particular, it does not provide direct access to the pole position and trajectory of the pole in the complex plane. Hence, we now transition to pole parameter extraction through the uniformized $\mathcal{S}$-matrix parameterization complemented by the model-dependent $\mathcal{K}$-matrix analysis.

\section{Pole Parameter Extraction}\label{sec:4}

In this section, we extract the pole position of the $T_{cc}^+$ by performing an extensive parametric fit to the LHCb invariant mass spectrum using the uniformized coupled-channel $\mathcal{S}$-matrix parameterization discussed in Section~\ref{sec:2}. To complement this, we also extract the pole parameters via a model-dependent approach using the two-channel $\mathcal{K}$-matrix formalism. Both fitting procedure is implemented using the \textsc{Minuit} minimization framework~\cite{James:1975} and is designed to determine the resonance mass and width from the analytic structure of the scattering amplitude.

From the minimal and robust classification results of the DNN workflow in Section~\ref{sec:3}, the fitting procedure only consider the minimal scenario of a single isolated pole located on the $[bt]$ sheet. The pole position is parameterized directly as
\begin{equation}
    \sqrt{s}^{(\text{pole})}_{T_{cc}^+} = M_{T_{cc}^+} - i \,\frac{\Gamma_{T_{cc}^+}}{2}
\end{equation}
where  $M_{T_{cc}^+}$ is the resonance mass and $\Gamma_{T_{cc}^+}$ is the resonance decay width. The predicted invariant mass spectrum is given by Eq.~\eqref{eq:dN/ds} and we modeled the non-resonant background as a smooth quadratic polynomial $b (\sqrt{s}) = c_0 + c_1\, \sqrt{s} + c_2\, s$ with coefficient $c_i$ determined in the fit.

To eliminate binning bias and  accurately capture the sharp threshold dynamics of the $T_{cc}^+$, the predicted yield is calculated by integrating the probability density function~\cite{Gligorov:2021sry} over the $0.5$ [MeV] bin width. In our numerical implementation, this is approximated via a multi-point grid average with a fine spacing of $0.005$ [MeV]. Crucially, the theoretical model is convolved with a double-Gaussian resolution kernel to match the LHCb detector resolution following their spectroscopic paper on $T_{cc}^+$~\cite{LHCb:2021auc}.\footnote{The double-Gaussian convolution kernel reads
 \begin{align*}
     \mathcal{R}_{\mathrm{LHCb}} (\sqrt{s}, \sqrt{s}_i) &= \kappa_{\mathrm{core}} \, \mathcal{G} (\sqrt{s}, \sqrt{s}_i, \sigma_\mathrm{core}) \nonumber \\
     & \qquad + (1-\kappa_{\mathrm{core}}) \, \mathcal{G} (\sqrt{s}, \sqrt{s}_i, \sigma_\mathrm{tail}).
 \end{align*}
 
 In particular, the core component is a narrow Gaussian with $\sigma_{\mathrm{core}} \approx 0.25$ [MeV], represents the primary detector resolution. The tail component is a broader Gaussian with $\sigma_{\mathrm{tail}} \approx 0.65$ [MeV] which captures the extended resolution effects. These components are combined with a core fraction $\kappa_{\mathrm{core}} = 0.70$, yielding a total effective root mean square (RMS) resolution of approximately $0.40$ [MeV]~\cite{LHCb:2021auc}.} The convolution is performed on a high-resolution grid extended by $5.0$ [MeV] to prevent edge artifacts. 

 The fit parameters are extracted via a multi-start $\chi^2$ cost function minimization using the \textsc{migrad} algorithm provided by \textsc{Minuit}~\cite{James:1975}. To mitigate sensitivity to local minima in the high-dimensional parameter space, we performed $10^3$ independent fits with randomized initial conditions within physically motivated parameter bounds.

\subsection{Uniformized \texorpdfstring{$\mathcal{S}$}{}-matrix}

\begin{figure*}[!ht]
    \centering
    
    \subfloat[\label{fig:minuitUSM}]{
        \includegraphics[width=0.48\linewidth]{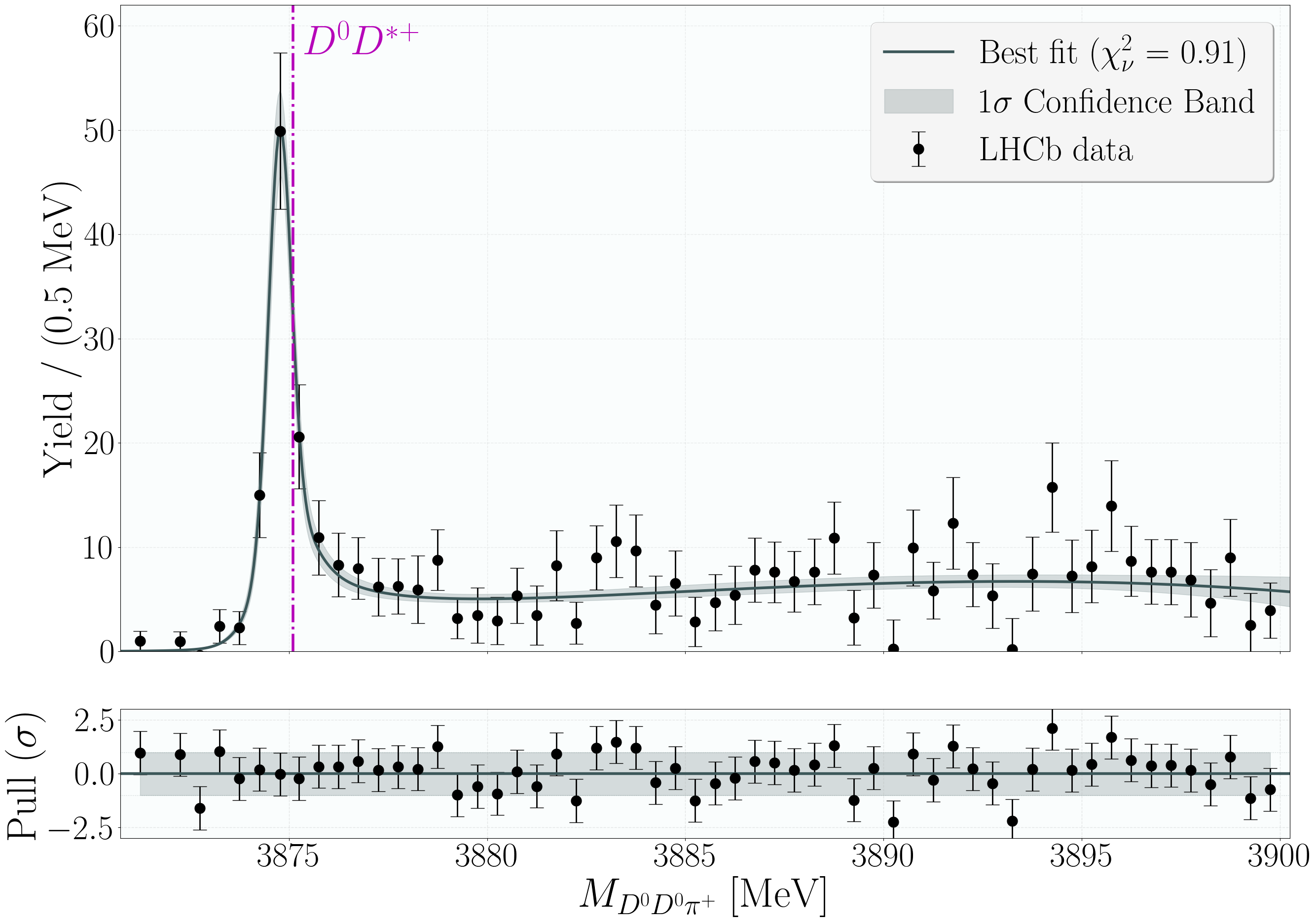}
    }
    \hfill
    \subfloat[\label{fig:minuitKM}]{
        \includegraphics[width=0.48\linewidth]{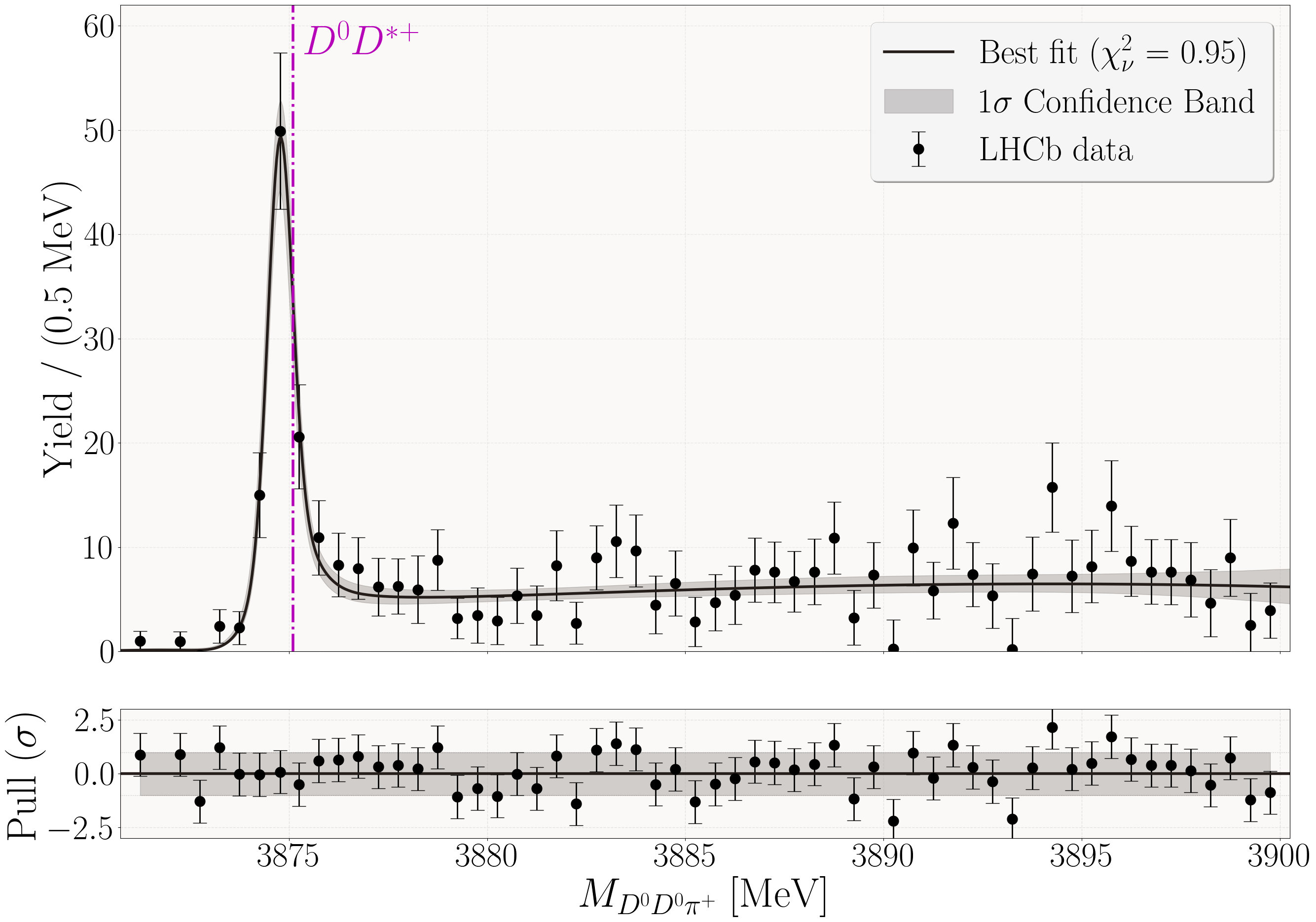}
    }
    
    \caption{\textbf{Best fit line shape and statistical diagnostics for the isolated pole in the $\left[bt\right]$ sheet model via  (a) the uniformized $\mathcal{S}$-matrix and (b) $\mathcal{K}$-matrix framework.}}
    \label{fig:main}
\end{figure*}

\begin{table*}[!ht]
    \centering
    \caption{\textbf{Extracted best fit pole parameters for an isolated pole on the \texorpdfstring{$[bt]$}{} sheet model.}}
    \setlength{\tabcolsep}{30pt}
    \renewcommand{\arraystretch}{1.5}
    \begin{tabular}{ccc}
    \hline \hline
        Pole Parameters & Uniformized $\mathcal{S}$-matrix & $\mathcal{K}$-matrix \\
        \hline
        $M_{T_{cc}^+}$ [MeV] & $3874.783 \pm 0.042$ & $3874.749 \pm 0.072$\\
        $\Gamma_{T_{cc}^+}$ [MeV] & $0.097 \pm 0.010$ & $0.232 \pm 0.031$\\
    \hline \hline
    \end{tabular}
    \label{tab:poleparams}
\end{table*}

To provide a model-independent baseline for the pole extraction, we first carry out a parametric fit via the uniformized $\mathcal{S}$-matrix. Here, the mass and width of the $T_{cc}^+$ are directly treated as fitting parameters. Figure~\ref{fig:minuitUSM} presents the invariant mass distribution of $D^0D^0\pi^+$ candidates with the best-fit result. The solid aqua green curve accurately reproduces the sharp, near-threshold enhancement corresponding to the $T_{cc}^+$ state, capturing the rapid rise at the $D^0D^{*+}$ threshold and the subsequent $D^0D^0\pi^+$ decay. The surrounding gray shaded region represents the $1\sigma$ confidence band, generated via Monte Carlo propagation of the fit parameter uncertainties. The confidence band is notably narrow, which indicates that the model parameters are constrained with high precision by the data.

The lower panel displays the pull distribution showing that the pulls are randomly distributed around zero without systematic structures, and the majority of data points fall within the expected $\pm 1 \sigma$ band (shaded gray rectangle). This statistical behavior, corroborated by the reduced chi-square value of $\chi_\nu^2 = 0.91$, confirms that an isolated pole in the $[bt]$ sheet model from the uniformized $\mathcal{S}$-matrix framework provides a robust description of the experimental spectrum.

\subsection{\texorpdfstring{$\mathcal{K}$}{}-matrix}

A complementary fit was also performed using a $\mathcal{K}$-matrix parameterization to extract the numerical values of the pole parameters. By construction, the $\mathcal{K}$-matrix formalism preserves the unitarity of the $\mathcal{S}$-matrix, thus ensuring conservation of probability~\cite{chung}. In the two-channel case relevant to the analysis, the first and second channels incorporated here are the $D^0 D^0 \pi^+$ and $D^0 D^{*+}$ channels. The $\mathcal{K}$-matrix is then constructed as a real symmetric $2 \times 2$ matrix:  

\begin{equation}
\label{Kconst}
\mathcal{K} = 
\begin{pmatrix}
\mathcal{K}_{11} & \mathcal{K}_{12} \\
\mathcal{K}_{21} & \mathcal{K}_{22}
\end{pmatrix},
\end{equation}
where the corresponding elements are treated as free parameters in the fitting scheme. In general, each element of the $\mathcal{K}$-matrix can be written as the sum of the resonant and non-resonant contributions,

\begin{equation}\label{eq:Kmat}
\mathcal{\hat{K}}_{ij} = \sum_{a} \frac{\gamma_{ai}\gamma_{aj}m_a \Gamma_a^0}{m_a^2 - m^2} + \hat{c}_{ij}.
\end{equation}

Here the $\gamma_{ai}$ denotes the coupling strength of resonance $a$ to the $i^{\text{th}}$ channel, while $m_a$ and $\Gamma_{a}^{0}$ represent the mass and width of the resonance. The term $\hat{c}_{ij}$ accounts for the non-resonant contribution.

We formalize and modify the general representation of the $\mathcal{K}$-matrix elements using the result of an isolated pole from Section~\ref{sec:3} as motivation. The presence of the isolated pole structure seen in our DNN result allows for the elimination of the resonance mass/width dependence in the usual $\mathcal{K}$-matrix element equation as seen in Eq.~\eqref{eq:Kmat}, effectively reducing the $\mathcal{K}$-matrix elements into energy-independent terms. This also reduces the number of free parameters present during the fitting scheme, allowing for a more accurate extraction of pole parameter values in a fitting perspective while also capturing the relevant resonance behavior near the peak and threshold.

The obtained $\mathcal{K}$-matrix parameters are used to build the scattering $T$-matrix using the relation defined by:

\begin{equation}
T = \rho^{1/2}\mathcal{K}(I-i\mathcal{K})^{-1}\rho^{1/2}
\end{equation}
where $I$ is the identity matrix and $\rho$ is the phase-space matrix which is a $2 \times 2$ diagonal matrix. The $T$-matrix is then used to reconstruct the invariant mass distribution using the differential event rate given by Eq.~\eqref{eq:dN/ds} that was introduced in Section~\ref{sec:3}.

We take advantage of the relation between the $\mathcal{S}$- and $T$-matrices mentioned in Section~\ref{sec:2} and use it together with the obtained best-fit values to numerically compute the poles of the $\mathcal{S}$-matrix, which correspond to peaks seen in the scattering data.  The resulting fit, as shown in Fig.~\ref{fig:minuitKM}, reproduces the narrow enhancement across the concerned energy region. The reduced chi-square value of the best fit is $\chi_{\nu}^2 = 0.95$, which suggests a good agreement between the model and data.

\subsection{Pole Position and Trajectory of \texorpdfstring{$T_{cc}^+$}{}}

In the fitting scheme, the global minimum was identified by the smallest $\chi^2$ value among all valid converged trials. Table~\ref{tab:poleparams} summarizes the best-fit pole parameters with their associated uncertainties. The extracted mass is remarkably consistent across both parameterizations, reflecting the clear threshold enhancement in the LHCb data. 

However, we observed a rather large deviation with the extracted width across both parametric framework. The discrepancy between the uniformized $\mathcal{S}$-matrix ($\Gamma_{T_{cc}^+} \sim 98$ [keV]) and $\mathcal{K}$-matrix ($\Gamma_{T_{cc}^+} \sim 232$ [keV]) highlights the inherent sensitivity of the $T_{cc}^+$ width to the choice of unitarization scheme. Given the detector resolution ($\sigma\approx 400$ [keV]) significantly exceeds both extracted widths, these two values define the envelope of our systematic uncertainty, consistently placing the $T_{cc}^+$ as a narrow, near-threshold state regardless of the specific mathematical framework employed.

Nonetheless, our result is physically consistent and sits between the LHCb's discovery paper~\cite{LHCb:2021vvq} which employed a Breit-Wigner parameterization extracting an apparent width of $\Gamma_{\mathrm{B.W.}} = 410.0 \pm 165.0$ [keV] and the rigorous spectroscopic LHCb analysis~\cite{LHCb:2021auc}, which employed a unitarized Breit-Wigner parameterization with momentum-dependent width, extracting a pole width $\Gamma_{\mathrm{U.B.W.}} = 48.0 \pm 2.0$ [keV].

We must point out however, that these extracted values should be viewed as effective pole parameters within a two-body, two-channel approximation. In particular, our model employs an effective two-body treatment of the $D^0D^0\pi^+$ system; Ref.~\cite{Du:2021zzh} argued that neglecting explicit three-body $DD\pi$ dynamics with its associated branch cut structure and the energy-dependent pion exchange dynamics inherently leads to an overestimation of the pole width. Moreover, our analysis is restricted to a two-channel framework, neglecting the $D^+ D^{*0}$ threshold. Since this third mass threshold lies only $\sim 1.4$ [MeV] above the $D^0D^{*+}$ channel, the omission likely forces the primary channels to absorb the spectral density that would otherwise be distributed across the isospin-multiplet, further deviating the extracted pole width.

In addition to the pole parameter extraction, we resolve the loosely-bound and virtual state ambiguity within the $[bt]$ pole topology. As mentioned in Section~\ref{sec:3}, an isolated pole in the $[bt]$ sheet cannot be definitively used to support the loosely-bound state interpretation for the $T_{cc}^+$ because such a pole topology can still be reached by a virtual state pole of the $D^0 D^{*+}$ channel due to strong interchannel coupling~\cite{Frazer:1964, PearceGibson, Fernandez-Ramirez:2019koa}. 

\begin{figure}[!ht]
    \centering
    \includegraphics[width=0.95\linewidth]{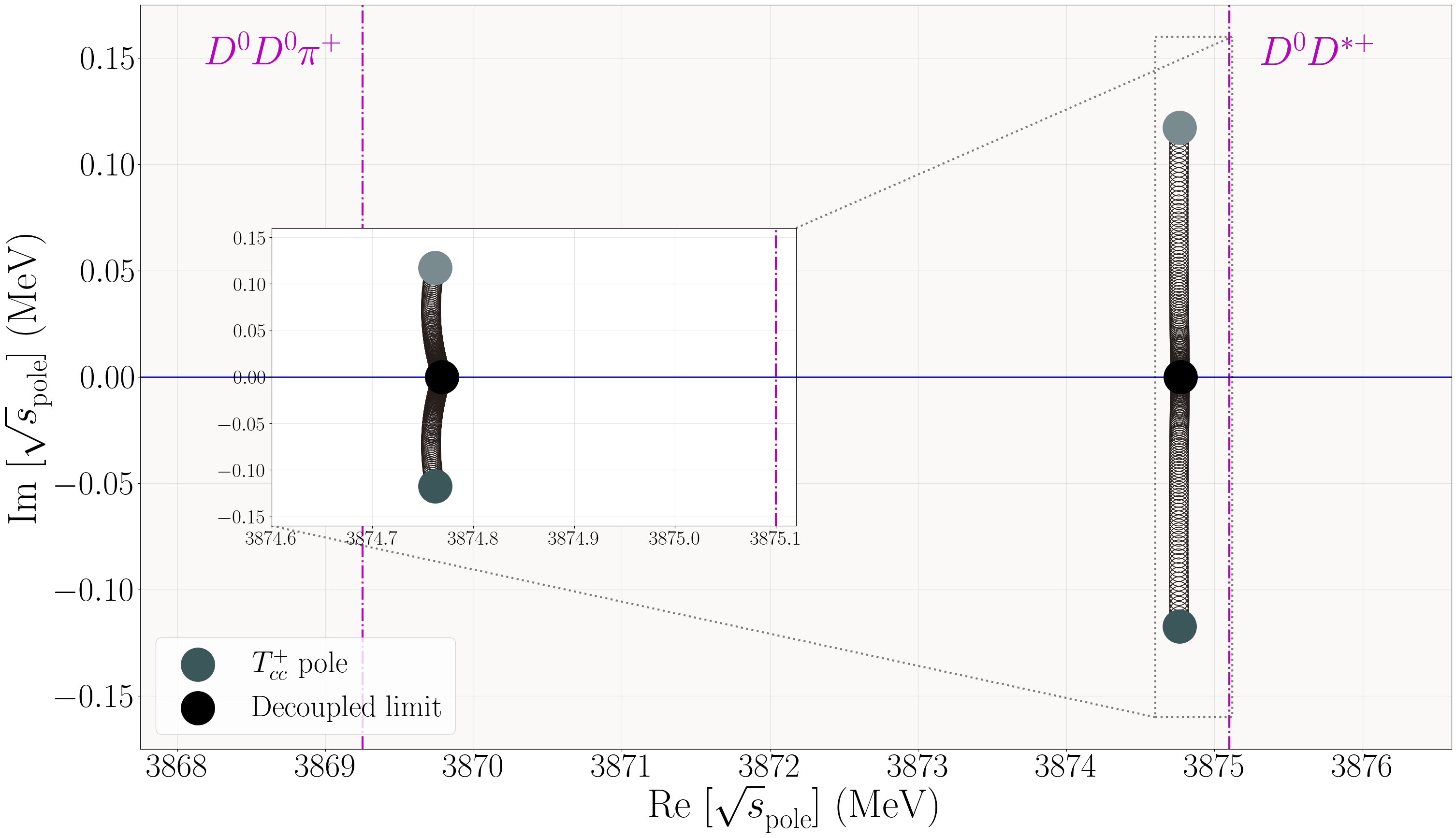}
\caption{\textbf{Trajectory of the pole found in the [$bt$] sheet}. Turning off the coupling to the $D^{0}D^{*+}$ channel shifts the pole towards the real energy axis, which is a signature of an unstable bound state.}
    \label{fig:kmatfittrajectory}
\end{figure}

To break this degeneracy, we trace the pole trajectory across the Riemann sheets by switching off the coupling to the second channel (Figure~\ref{fig:kmatfittrajectory}). Within our $\mathcal{K}$-matrix formalism, this is equivalent to setting the off-diagonal elements to zero to indicate the type of the resulting one-channel pole. In this case, the pole moves toward the real energy axis below the $D^{0}D^{*+}$ threshold. Using the pole-counting argument, the observed pole in the [$bt$] sheet is classified as a loosely-bound state of the $D^0D^{*+}$ system.

Ultimately, while the extracted pole parameters for the $T_{cc}^+$ are subject to the inherent limitations of a simplified two-body, two-channel parameterization, this analysis serves as a successful proof-of-concept for our broader methodology. We have demonstrated that a \textit{two-pronged strategy} (utilizing DNNs for topological model selection followed by multi-trial \textsc{Minuit}-based pole extraction) provides a robust framework for identifying the nature of exotic hadron candidates.

The simplifications employed within the uniformized $\mathcal{S}$-matrix model were tactical necessities. In the present work, we utilized a two-channel uniformized $\mathcal{S}$-matrix framework~\cite{Kato:1965} to leverage its capacity for independent and controlled pole position formulation~\cite{Santos:2023gfh} for a rigorous inference-assisted pole topology DNN classification. While alternative parameterizations, such as the $\mathcal{K}$-matrix, could in principle accommodate more channels or three-body effects, they lack direct mapping between $\mathcal{S}$-matrix analyticity and pole locations required to generate the high-fidelity, labeled training datasets essential for our data-driven approach.

\section{Conclusions}\label{sec:conclusion}

This work implemented a two-pronged strategy combining inference-assisted pole topology classification and parametric fitting to extract pole parameters to analyze the doubly-charmed tetraquark candidate $T_{cc}^+$. A hierarchical DNN classification scheme overcame intrinsic degeneracies among line shapes of viable pole topologies. While lhc effects introduce additional analytic structures and moderately reduce classification performance, the dominant pole topology remains robust: DNN classifiers trained on amplitudes with and without lhc effects consistently feature an isolated pole on the $[bt]$ sheet. Quantitative pole parameter extraction demonstrate that while the mass of the $T_{cc}^+$ remains remarkably stable across disparate parametric schemes, the observed sensitivity in the pole width highlights systematic trade-offs of a simplified coupled-channel framework. Finally, pole trajectory analysis supports a quasi-bound $D^0D^{*+}$ molecular interpretation for the $T_{cc}^+$.

 Moving forward, regression-based deep neural networks offer a promising complementary avenue for pole parameter determination, enabling a direct, data-driven mapping from line shapes to pole positions beyond conventional parametric fitting schemes.

\begin{acknowledgments}
JBP acknowledges the scholarship support through DOST-ASTHRDP. KTRC acknowledges the support from the Pisayian graduate scholarship.
\end{acknowledgments}

\bibliography{TccML}

\onecolumngrid
\newpage
% \appendix
\section*{Appendix}
\label{sec:appendix}

\setcounter{equation}{0}
\setcounter{figure}{0}
\setcounter{table}{0}
\renewcommand{\theequation}{A.\arabic{equation}}
\renewcommand{\thefigure}{A.\arabic{figure}}
\renewcommand{\thetable}{A.\arabic{table}}

\subsection{Pole-topology classes in a coupled two-channel system}

We define 35 distinct pole topology classes, each characterized by a triplet $[N_{bt}, N_{bb}, N_{tb}]$ with $N_i \in [0,4]$ counting poles on the corresponding Riemann sheet. Poles on the $[tt]$ sheet are excluded to satisfy causality~\cite{Screaton:1969he, Minerbo:1971gg}. This basis captures the possible analytic structures of coupled-channel amplitudes, where poles migrate between sheets under varying coupling and background conditions~\cite{PearceGibson, Morgan:1992}. The complete set spans pole-free amplitudes to configurations with multiple nearby poles, ensuring that no analytically allowed topology is excluded \emph{a priori} in the subsequent DNN analysis.

\begin{table*}[!ht]
\centering
\caption{\textbf{Pole-topology classification in a coupled two-channel system.}
The 35 configurations ($00$–$34$) are labeled by $[N_{bt},N_{bb},N_{tb}]$, indicating pole multiplicities on the $[bt]$, $[bb]$, and $[tb]$ sheets. They are organized into seven structural families (SF1–SF7) based on shared analytic structure.}
\label{tab:pole_class}
\begin{tabular}{ccl}
\hline \hline
\textbf{Label} & \textbf{Pole Structure} & \textbf{Physical Interpretation} \\
\hline \hline
\multicolumn{3}{l}{\textbf{SF1}} \\
\hline
00 & [$0_{bt}, 0_{bb}, 0_{tb}$] & pure non-resonant background\\
01 & [$1_{bt}, 0_{bb}, 0_{tb}$] & hadronic molecule (loosely-bound state or a virtual state)~\cite{Morgan:1992}\\
02 & [$0_{bt}, 1_{bb}, 0_{tb}$] & not possible for ERE~\cite{Frazer:1964}\\
03 & [$0_{bt}, 0_{bb}, 1_{tb}$] & virtual state\\
04 & [$2_{bt}, 0_{bb}, 0_{tb}$] & 2 molecular states [commonly associated with $KN$ final state]\\
\hline
\multicolumn{3}{l}{\textbf{SF2}} \\
\hline
05 & [$0_{bt}, 2_{bb}, 0_{tb}$] & --\\
06 & [$0_{bt}, 0_{bb}, 2_{tb}$] & --\\
07 & [$1_{bt}, 1_{bb}, 0_{tb}$] & may be interpreted as a compact state coupled to the lower channel\\
08 & [$1_{bt}, 0_{bb}, 1_{tb}$] & hadronic molecule + virtual state\\
09 & [$0_{bt}, 1_{bb}, 1_{tb}$] & may be interpreted as a compact state coupled to the higher channel\\
\hline
\multicolumn{3}{l}{\textbf{SF3}} \\
\hline
10 & [$3_{bt}, 0_{bb}, 0_{tb}$] & --\\
11 & [$0_{bt}, 3_{bb}, 0_{tb}$] & --\\
12 & [$0_{bt}, 0_{bb}, 3_{tb}$] & --\\
13 & [$2_{bt}, 1_{bb}, 0_{tb}$] & hadronic molecule + compact state \\
14 & [$2_{bt}, 0_{bb}, 1_{tb}$] & --\\
\hline
\multicolumn{3}{l}{\textbf{SF4}} \\
\hline
15 & [$1_{bt}, 2_{bb}, 0_{tb}$] & --\\
16 & [$0_{bt}, 2_{bb}, 1_{tb}$] & --\\
17 & [$1_{bt}, 0_{bb}, 2_{tb}$] & --\\
18 & [$0_{bt}, 1_{bb}, 2_{tb}$] & compact state coupled to the higher channel + virtual state\\
19 & [$1_{bt}, 1_{bb}, 1_{tb}$] & compact state coupled to the lower channel + virtual state~\cite{Santos:2024bqr}\\
\hline
\multicolumn{3}{l}{\textbf{SF5}} \\
\hline
20 & [$4_{bt}, 0_{bb}, 0_{tb}$] & --\\
21 & [$0_{bt}, 4_{bb}, 0_{tb}$] & --\\
22 & [$0_{bt}, 0_{bb}, 4_{tb}$] & --\\
23 & [$3_{bt}, 1_{bb}, 0_{tb}$] & --\\
24 & [$3_{bt}, 0_{bb}, 1_{tb}$] & --\\
\hline
\multicolumn{3}{l}{\textbf{SF6}} \\
\hline
25 & [$1_{bt}, 3_{bb}, 0_{tb}$] & --\\
26 & [$0_{bt}, 3_{bb}, 1_{tb}$] & --\\
27 & [$1_{bt}, 0_{bb}, 3_{tb}$] & --\\
28 & [$0_{bt}, 1_{bb}, 3_{tb}$] & --\\
29 & [$2_{bt}, 2_{bb}, 0_{tb}$] & possible two compact states\\
\hline
\multicolumn{3}{l}{\textbf{SF7}} \\
\hline
30 & [$2_{bt}, 0_{bb}, 2_{tb}$] & --\\
31 & [$0_{bt}, 2_{bb}, 2_{tb}$] & --\\
32 & [$2_{bt}, 1_{bb}, 1_{tb}$] & --\\
33 & [$1_{bt}, 2_{bb}, 1_{tb}$] & compact state + virtual state pole with non-vanishing width~\cite{Frohnert:2025usi}\\
34 & [$1_{bt}, 1_{bb}, 2_{tb}$] & --\\
\hline \hline
\end{tabular}
\end{table*}

\newpage
\subsection{Summary of DNN-Based Pole Topology Inference}

The original $35$ pole-topology configurations were grouped into seven structural families according to pole multiplicity and Riemann-sheet distribution. Ensemble of DNN classifiers were trained for each family and inference was performed on pseudo-empirical amplitudes generated from the LHCb line shape. 

For amplitudes without lhc effects, dominant pole topology emerged in each structural family. The leading candidates were class $01: [1_{bt}, 0_{bb}, 0_{tb}]$, class $07: [1_{bt}, 1_{bb}, 0_{tb}]$, class $10: [3_{bt}, 0_{bb}, 0_{tb}]$, class $19: [1_{bt}, 1_{bb}, 1_{tb}]$, class $22: [0_{bt}, 0_{bb}, 4_{tb}]$, class $27/28: [1_{bt}, 0_{bb}, 3_{tb}]/[0_{bt}, 1_{bb}, 3_{tb}]$, and class $33: [1_{bt}, 2_{bb}, 1_{tb}]$.

\begin{figure}[!ht]
    \centering
    \includegraphics[width=0.99\linewidth]{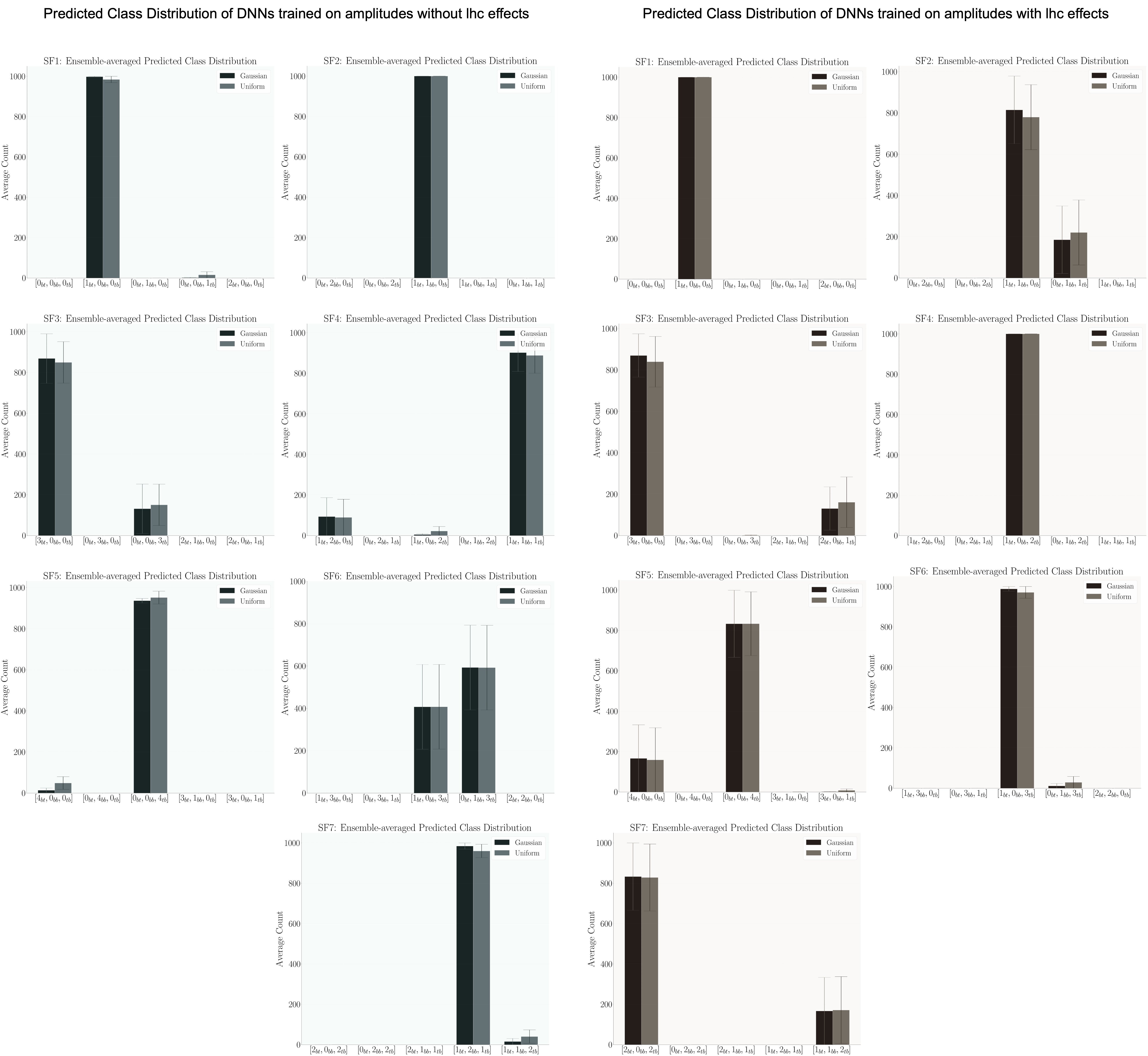}
    \caption{Ensemble mean with standard error of the predicted class distribution for the LHCb data for every structural families (SF1-SF7).}
    \label{fig:SFPCD}
\end{figure}
Including lhc effects reduced the viable set to structurally stable configurations, primarily classes $01: [1_{bt}, 0_{bb}, 0_{tb}]$,  $07: [1_{bt}, 1_{bb}, 0_{tb}]$, $10: [3_{bt}, 0_{bb}, 0_{tb}]$, and $22: [0_{bt}, 0_{bb}, 4_{tb}]$. A subsequent grouped finalist selection consistently identified classes $01: [1_{bt}, 0_{bb}, 0_{tb}]$ and $07: [1_{bt}, 1_{bb}, 0_{tb}]$ as the two most dominant hypotheses across all considered analytic scenarios. These configurations were therefore carried forward to the focused binary classification analysis and parametric pole extraction presented in the main text.
\begin{figure}[!ht]
    \centering
    \includegraphics[width=0.99\linewidth]{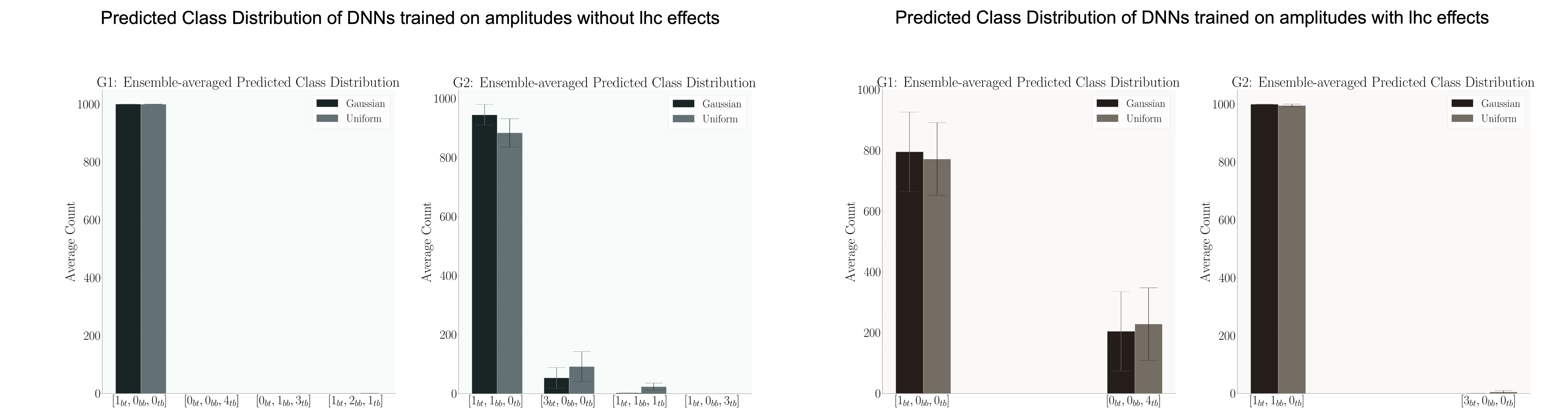}
    \caption{Ensemble mean with standard error of the predicted class distribution for the LHCb data for two groups of candidate classes .}
    \label{fig:GPCD}
\end{figure}

\subsection{Fitting parameters}

\begin{table*}[!ht]
    \centering
    \begin{threeparttable}
        \caption{\textbf{Extracted best fit parameters for an isolated pole on the \texorpdfstring{$[bt]$}{} sheet model.}}
        \setlength{\tabcolsep}{20pt}
        \renewcommand{\arraystretch}{1.4}
        \begin{tabular}{ccc}
            \hline \hline
            Parameter & {Uniformized $\mathcal{S}$-matrix} & {$\mathcal{K}$-matrix} \\
            \hline
            $M_{T_{cc}^+}$ [MeV]      & $3874.78 \pm 0.04$ & ${-}$ \\
            $\Gamma_{T_{cc}^+}$ [keV] & $97.20 \pm 9.40$   & ${-}$ \\
            $\mathcal{K}_{11}$        & ${-}$            & $-5.84 \pm 0.74$ \\
            $\mathcal{K}_{12}$        & ${-}$            & $-15.93 \pm 2.66$ \\
            $\mathcal{K}_{22}$        & ${-}$            & $-76.57 \pm 7.11$ \\
            $\mathcal{A}$\tnote{*}    & $1.00 \pm 0.00$    & $1.00 \pm 0.00$ \\
            $\alpha_1$\tnote{*}       & $1.68 \pm 0.17$    & $52.02 \pm 0.52$ \\
            $\alpha_2$\tnote{**}        & $1.00 \pm 0.00$    & $1.00 \pm 0.00$ \\
            $c_0$\tnote{**}             & $0.00 \pm 0.00$    & $0.00 \pm 0.00$ \\
            $c_1$                     & $(2.78 \pm 0.69) \times 10^{-3}$      & $(-0.72 \pm 0.50) \times 10^1$ \\
            $c_2$                     & $(-8.30 \pm 0.33) \times 10^{-5}$     & $(-0.18 \pm 1.06) \times 10^{-1}$ \\
            \hline \hline
        \end{tabular}
        
        \begin{tablenotes}
            \small
            \item[*] Fixed to its best fit values from prior runs.
            \item[**] Fixed
        \end{tablenotes}
    \end{threeparttable}
    \label{tab:fitparams}
\end{table*}

\newpage
\begin{table*}[!ht]
    \centering
    \caption{Comprehensive performance summary of the $50$ DNN models trained on amplitudes without lhc effects. The table includes the model architecture, validation accuracy, and the ensemble-averaged predicted class distribution from inference on LHCb data using both Gaussian and Uniform sampling. The final row shows the ensemble mean for the inference counts.}
    \begin{tabular}{ccccccc}
    \hline \hline
DNN & Architecture & Accuracy & \multicolumn{2}{c}{Gaussian Sampling} & \multicolumn{2}{c}{Uniform Sampling} \\
\cline{4-7}
    &              &          & $[1_{bt}, 0_{bb}, 0_{tb}]$ & $[1_{bt}, 1_{bb}, 0_{tb}]$ & $[1_{bt}, 0_{bb}, 0_{tb}]$ & $[1_{bt}, 1_{bb}, 0_{tb}]$ \\
\hline
        1 & [$200, 200$] & $93\%$ & $1000.0$ & $0.0$ & $1000.0$ & $0.0$ \\
        2 & [$300, 300$] & $94\%$ & $1000.0$ & $0.0$ & $1000.0$ & $0.0$ \\
        3 & [$400, 400$] & $94\%$ & $1000.0$ & $0.0$ & $1000.0$ & $0.0$ \\
        4 & [$500, 500$] & $95\%$ & $1000.0$ & $0.0$ & $997.0$ & $3.0$ \\
        5 & [$600, 600$] & $94\%$ & $1000.0$ & $0.0$ & $999.0$ & $1.0$ \\
        6 & [$700, 700$] & $94\%$ & $1000.0$ & $0.0$ & $1000.0$ & $0.0$ \\
        7 & [$800, 800$] & $94\%$ & $996.0$ & $4.0$ & $927.0$ & $73.0$ \\
        8 & [$900, 900$] & $95\%$ & $1000.0$ & $0.0$ & $1000.0$ & $0.0$ \\
        9 & [$200, 200, 200$] & $95\%$ & $1000.0$ & $0.0$ & $1000.0$ & $0.0$ \\
        10 & [$200, 300, 200$] & $95\%$ & $1000.0$ & $0.0$ & $1000.0$ & $0.0$ \\
        11 & [$300, 300, 300$] & $95\%$ & $1000.0$ & $0.0$ & $999.0$ & $1.0$ \\
        12 & [$300, 400, 300$] & $95\%$ & $1000.0$ & $0.0$ & $1000.0$ & $0.0$ \\
        13 & [$400, 400, 400$] & $95\%$ & $1000.0$ & $0.0$ & $1000.0$ & $0.0$ \\
        14 & [$400, 500, 400$] & $95\%$ & $1000.0$ & $0.0$ & $999.0$ & $1.0$ \\
        15 & [$500, 500, 500$] & $95\%$ & $982.0$ & $18.0$ & $872.0$ & $128.0$ \\
        16 & [$500, 600, 500$] & $94\%$ & $1000.0$ & $0.0$ & $1000.0$ & $0.0$ \\
        17 & [$600, 600, 600$] & $94\%$ & $1000.0$ & $0.0$ & $1000.0$ & $0.0$ \\
        18 & [$600, 700, 600$] & $95\%$ & $1000.0$ & $0.0$ & $1000.0$ & $0.0$ \\
        19 & [$700, 700, 700$] & $95\%$ & $1000.0$ & $0.0$ & $985.0$ & $15.0$ \\
        20 & [$700, 800, 700$] & $94\%$ & $1000.0$ & $0.0$ & $997.0$ & $3.0$ \\
        21 & [$800, 800, 800$] & $95\%$ & $1000.0$ & $0.0$ & $983.0$ & $17.0$ \\
        22 & [$200, 200, 200, 200$] & $94\%$ & $1000.0$ & $0.0$ & $1000.0$ & $0.0$ \\
        23 & [$200, 300, 300, 200$] & $94\%$ & $1000.0$ & $0.0$ & $998.0$ & $2.0$ \\
        24 & [$300, 300, 300, 300$] & $94\%$ & $1000.0$ & $0.0$ & $999.0$ & $1.0$ \\
        25 & [$300, 400, 400, 300$] & $95\%$ & $1000.0$ & $0.0$ & $1000.0$ & $0.0$ \\
        26 & [$400, 400, 400, 400$] & $95\%$ & $1000.0$ & $0.0$ & $1000.0$ & $0.0$ \\
        27 & [$400, 500, 500, 400$] & $95\%$ & $1000.0$ & $0.0$ & $1000.0$ & $0.0$ \\
        28 & [$500, 500, 500, 500$] & $95\%$ & $1000.0$ & $0.0$ & $1000.0$ & $0.0$ \\
        29 & [$500, 600, 600, 500$] & $94\%$ & $1000.0$ & $0.0$ & $998.0$ & $2.0$ \\
        30 & [$600, 600, 600, 600$] & $94\%$ & $1000.0$ & $0.0$ & $1000.0$ & $0.0$ \\
        31 & [$600, 700, 700, 600$] & $94\%$ & $1000.0$ & $0.0$ & $1000.0$ & $0.0$ \\
        32 & [$700, 700, 700, 700$] & $95\%$ & $993.0$ & $7.0$ & $948.0$ & $52.0$ \\
        33 & [$700, 800, 800, 700$] & $94\%$ & $1000.0$ & $0.0$ & $1000.0$ & $0.0$ \\
        34 & [$800, 800, 800, 800$] & $94\%$ & $1000.0$ & $0.0$ & $1000.0$ & $0.0$ \\
        35 & [$200, 200, 200, 200, 200$] & $95\%$ & $1000.0$ & $0.0$ & $1000.0$ & $0.0$ \\
        36 & [$200, 300, 300, 300, 200$] & $95\%$ & $937.0$ & $63.0$ & $809.0$ & $191.0$ \\
        37 & [$200, 300, 400, 300, 200$] & $95\%$ & $1000.0$ & $0.0$ & $1000.0$ & $0.0$ \\
        38 & [$300, 300, 300, 300, 300$] & $95\%$ & $1000.0$ & $0.0$ & $999.0$ & $1.0$ \\
        39 & [$300, 400, 400, 400, 300$] & $95\%$ & $1000.0$ & $0.0$ & $1000.0$ & $0.0$ \\
        40 & [$300, 400, 500, 400, 300$] & $95\%$ & $1000.0$ & $0.0$ & $998.0$ & $2.0$ \\
        41 & [$400, 400, 400, 400, 400$] & $95\%$ & $1000.0$ & $0.0$ & $974.0$ & $26.0$ \\
        42 & [$400, 500, 500, 500, 400$] & $95\%$ & $990.0$ & $10.0$ & $927.0$ & $73.0$ \\
        43 & [$400, 500, 600, 500, 400$] & $94\%$ & $1000.0$ & $0.0$ & $999.0$ & $1.0$ \\
        44 & [$500, 500, 500, 500, 500$] & $95\%$ & $1000.0$ & $0.0$ & $1000.0$ & $0.0$ \\
        45 & [$500, 600, 600, 600, 500$] & $95\%$ & $1000.0$ & $0.0$ & $997.0$ & $3.0$ \\
        46 & [$500, 600, 700, 600, 500$] & $95\%$ & $1000.0$ & $0.0$ & $1000.0$ & $0.0$ \\
        47 & [$600, 600, 600, 600, 600$] & $95\%$ & $1000.0$ & $0.0$ & $1000.0$ & $0.0$ \\
        48 & [$600, 700, 700, 700, 600$] & $95\%$ & $1000.0$ & $0.0$ & $1000.0$ & $0.0$ \\
        49 & [$600, 700, 800, 700, 600$] & $95\%$ & $1000.0$ & $0.0$ & $995.0$ & $5.0$ \\
        50 & [$700, 700, 700, 700, 700$] & $94\%$ & $1000.0$ & $0.0$ & $1000.0$ & $0.0$ \\
        \hline
         &  &  & $997.96 \pm 1.32$ & $2.04 \pm 1.32$ & $987.98 \pm 4.96$ & $12.02 \pm 4.96$ \\
        \hline \hline
    \end{tabular}
    \label{tab:DNN_summary}
\end{table*}

\begin{table*}[!ht]
    \centering
    \caption{Comprehensive performance summary of the 50 DNN models trained on amplitudes with lhc effects from two-pion exchange. The table includes the model architecture, validation accuracy, and the ensemble-averaged predicted class distribution from inference on LHCb data using both Gaussian and Uniform sampling. The final row shows the ensemble mean for the inference counts.}
    \begin{tabular}{ccccccc}
    \hline \hline
DNN & Architecture & Accuracy & \multicolumn{2}{c}{Gaussian Sampling} & \multicolumn{2}{c}{Uniform Sampling} \\
\cline{4-7}
    &              &          & $[1_{bt}, 0_{bb}, 0_{tb}]$ & $[1_{bt}, 1_{bb}, 0_{tb}]$ & $[1_{bt}, 0_{bb}, 0_{tb}]$ & $[1_{bt}, 1_{bb}, 0_{tb}]$ \\
\hline  
        1 & [$200, 200$] & $84\%$ & $1000.0$ & $0.0$ & $1000.0$ & $0.0$ \\
        2 & [$300, 300$] & $85\%$ & $1000.0$ & $0.0$ & $1000.0$ & $0.0$ \\
        3 & [$400, 400$] & $83\%$ & $1000.0$ & $0.0$ & $1000.0$ & $0.0$ \\
        4 & [$500, 500$] & $85\%$ & $1000.0$ & $0.0$ & $1000.0$ & $0.0$ \\
        5 & [$600, 600$] & $85\%$ & $1000.0$ & $0.0$ & $1000.0$ & $0.0$ \\
        6 & [$700, 700$] & $83\%$ & $1000.0$ & $0.0$ & $1000.0$ & $0.0$ \\
        7 & [$800, 800$] & $83\%$ & $1000.0$ & $0.0$ & $1000.0$ & $0.0$ \\
        8 & [$900, 900$] & $81\%$ & $1000.0$ & $0.0$ & $1000.0$ & $0.0$ \\
        9 & [$200, 200, 200$] & $87\%$ & $1000.0$ & $0.0$ & $1000.0$ & $0.0$ \\
        10 & [$200, 300, 200$] & $86\%$ & $1000.0$ & $0.0$ & $1000.0$ & $0.0$ \\
        11 & [$300, 300, 300$] & $86\%$ & $1000.0$ & $0.0$ & $1000.0$ & $0.0$ \\
        12 & [$300, 400, 300$] & $86\%$ & $1000.0$ & $0.0$ & $1000.0$ & $0.0$ \\
        13 & [$400, 400, 400$] & $85\%$ & $1000.0$ & $0.0$ & $994.0$ & $0.0$ \\
        14 & [$400, 500, 400$] & $86\%$ & $1000.0$ & $0.0$ & $998.0$ & $2.0$ \\
        15 & [$500, 500, 500$] & $87\%$ & $1000.0$ & $0.0$ & $999.0$ & $1.0$ \\
        16 & [$500, 600, 500$] & $84\%$ & $549.0$ & $451.0$ & $539.0$ & $461.0$ \\
        17 & [$600, 600, 600$] & $84\%$ & $1000.0$ & $0.0$ & $1000.0$ & $0.0$ \\
        18 & [$600, 700, 600$] & $85\%$ & $1000.0$ & $0.0$ & $1000.0$ & $0.0$ \\
        19 & [$700, 700, 700$] & $85\%$ & $1000.0$ & $0.0$ & $1000.0$ & $0.0$ \\
        20 & [$700, 800, 700$] & $86\%$ & $1000.0$ & $0.0$ & $1000.0$ & $0.0$ \\
        21 & [$800, 800, 800$] & $86\%$ & $1000.0$ & $0.0$ & $991.0$ & $9.0$ \\
        22 & [$200, 200, 200, 200$] & $86\%$ & $1000.0$ & $0.0$ & $1000.0$ & $0.0$ \\
        23 & [$200, 300, 300, 200$] & $86\%$ & $1000.0$ & $0.0$ & $987.0$ & $13.0$ \\
        24 & [$300, 300, 300, 300$] & $85\%$ & $990.0$ & $10.0$ & $922.0$ & $78.0$ \\
        25 & [$300, 400, 400, 300$] & $85\%$ & $1000.0$ & $0.0$ & $1000.0$ & $0.0$ \\
        26 & [$400, 400, 400, 400$] & $86\%$ & $1000.0$ & $0.0$ & $1000.0$ & $0.0$ \\
        27 & [$400, 500, 500, 400$] & $85\%$ & $1000.0$ & $0.0$ & $995.0$ & $5.0$ \\
        28 & [$500, 500, 500, 500$] & $86\%$ & $1000.0$ & $0.0$ & $1000.0$ & $0.0$ \\
        29 & [$500, 600, 600, 500$] & $86\%$ & $1000.0$ & $0.0$ & $1000.0$ & $0.0$ \\
        30 & [$600, 600, 600, 600$] & $84\%$ & $1000.0$ & $0.0$ & $1000.0$ & $0.0$ \\
        31 & [$600, 700, 700, 600$] & $85\%$ & $1000.0$ & $0.0$ & $1000.0$ & $0.0$ \\
        32 & [$700, 700, 700, 700$] & $83\%$ & $1000.0$ & $0.0$ & $1000.0$ & $0.0$ \\
        33 & [$700, 800, 800, 700$] & $84\%$ & $1000.0$ & $0.0$ & $1000.0$ & $0.0$ \\
        34 & [$800, 800, 800, 800$] & $84\%$ & $1000.0$ & $0.0$ & $1000.0$ & $0.0$ \\
        35 & [$200, 200, 200, 200, 200$] & $86\%$ & $1000.0$ & $0.0$ & $1000.0$ & $0.0$ \\
        36 & [$200, 300, 300, 300, 200$] & $86\%$ & $1000.0$ & $0.0$ & $1000.0$ & $0.0$ \\
        37 & [$200, 300, 400, 300, 200$] & $85\%$ & $1000.0$ & $0.0$ & $1000.0$ & $0.0$ \\
        38 & [$300, 300, 300, 300, 300$] & $83\%$ & $1000.0$ & $0.0$ & $1000.0$ & $0.0$ \\
        39 & [$300, 400, 400, 400, 300$] & $85\%$ & $1000.0$ & $0.0$ & $1000.0$ & $0.0$ \\
        40 & [$300, 400, 500, 400, 300$] & $84\%$ & $1000.0$ & $0.0$ & $1000.0$ & $0.0$ \\
        41 & [$400, 400, 400, 400, 400$] & $85\%$ & $1000.0$ & $0.0$ & $1000.0$ & $0.0$ \\
        42 & [$400, 500, 500, 500, 400$] & $85\%$ & $1000.0$ & $0.0$ & $1000.0$ & $0.0$ \\
        43 & [$400, 500, 600, 500, 400$] & $84\%$ & $1000.0$ & $0.0$ & $1000.0$ & $0.0$ \\
        44 & [$500, 500, 500, 500, 500$] & $82\%$ & $1000.0$ & $0.0$ & $1000.0$ & $0.0$ \\
        45 & [$500, 600, 600, 600, 500$] & $86\%$ & $923.0$ & $77.0$ & $824.0$ & $176.0$ \\
        46 & [$500, 600, 700, 600, 500$] & $85\%$ & $1000.0$ & $0.0$ & $1000.0$ & $0.0$ \\
        47 & [$600, 600, 600, 600, 600$] & $83\%$ & $1000.0$ & $0.0$ & $1000.0$ & $0.0$ \\
        48 & [$600, 700, 700, 700, 600$] & $84\%$ & $1000.0$ & $0.0$ & $1000.0$ & $0.0$ \\
        49 & [$600, 700, 800, 700, 600$] & $84\%$ & $1000.0$ & $0.0$ & $1000.0$ & $0.0$ \\
        50 & [$700, 700, 700, 700, 700$] & $81\%$ & $1000.0$ & $0.0$ & $1000.0$ & $0.0$ \\
        \hline
         &  &  & $989.24 \pm 9.12$ & $10.76 \pm 9.12$ & $984.98 \pm 9.87$ & $15.02 \pm 9.87$ \\
        \hline \hline
    \end{tabular}
    \label{tab:DNN_summary_LHC}
\end{table*}

\end{document}